\documentclass[letterpaper, doublecolumn]{IEEEtran}
\usepackage{graphicx,epsfig,bm}
\usepackage{color}

\usepackage{amsmath,epsfig} % easier math formulae, align, subequation
\usepackage{enumerate}
\usepackage{amsfonts}
\usepackage{bm}
\usepackage{amssymb}
\usepackage{algorithm,algcompatible}
\usepackage{float}
\usepackage[english]{babel}
\usepackage{cite}
\usepackage{dsfont}
\usepackage{epstopdf}
\usepackage{color}

\algnewcommand\INPUT{\item[\textbf{Input:}]}%
\algnewcommand\OUTPUT{\item[\textbf{Output:}]}%

\newtheorem{thm}{Theorem}
\newtheorem{cor}{Corollary}

\newtheorem{prop}{Proposition}

\long\def\symbolfootnote[#1]#2{\begingroup%
\def\thefootnote{\fnsymbol{footnote}}\footnote[#1]{#2}\endgroup}

\newcommand{\beq}{\begin{equation}}
\newcommand{\eeq}{\end{equation}}
\newcommand{\beqa}{\begin{eqnarray}}
\newcommand{\eeqa}{\end{eqnarray}}

\newcommand{\pr}{{p}}

\newcommand{\bmlambda}{\bm{\lambda}}

\newcommand{\matD}{\mathbf{W}}

\newcommand{\matH}{\mathbf{H}}

\newcommand{\matI}{\mathbf{I}}

\newcommand{\Ex}{\mathbb{E}}

\newcommand{\bmLambda}{\bm{\Lambda}}

\newcommand{\expectation}{\ensuremath{\mathbb{E}}}
\newcommand{\Expt}{\expectation}
\newcommand{\probability}{\ensuremath{\mathbb{P}}}
\newcommand{\Prob}{\probability}

\allowdisplaybreaks[1]

\makeatletter
\newcommand{\vast}{\bBigg@{3}}
\newcommand{\Vast}{\bBigg@{4}}
\makeatother
\newcommand{\squeezeup}{\vspace{-2.5mm}}

\usepackage{xr}
\usepackage{zref-base}
\usepackage{atveryend}
\makeatletter
\AfterLastShipout{%
  \if@filesw   
    \begingroup
      \zref@setcurrent{default}{\the\value{equation}}%
      \zref@wrapper@immediate{%
        \zref@labelbyprops{eq-last}{default}%
      }%
    \endgroup
  \fi
}
\makeatother

\title{Lower Bounds on the Bayes Risk of the Bayesian BTL Model with Applications to Comparison Graphs}
\author{Mine~Alsan, \IEEEmembership{Member,~IEEE}, Ranjitha~Prasad and~Vincent~Y.~F.~Tan, \IEEEmembership{Senior~Member,~IEEE} 
\thanks{The authors are with the Department of Electrical and Computer Engineering, National University of Singapore (NUS) (emails: elemine@nus.edu.sg, ranjitha.p@gmail.com, vtan@nus.edu.sg). The third author is also with the Department of Mathematics, NUS.
 } }

\begin{document}

\maketitle
\begin{abstract}
We consider the problem of aggregating pairwise comparisons to obtain a consensus ranking order over a collection of objects. We use the popular Bradley-Terry-Luce (BTL) model which allows us to probabilistically describe pairwise comparisons between objects. In particular, we employ the Bayesian BTL model which allows for meaningful prior assumptions and to cope with situations where the number of objects is large and the number of comparisons between some objects is small or even zero. For the conventional Bayesian BTL model, we derive information-theoretic lower bounds on the Bayes risk of estimators for norm-based distortion functions. We compare the information-theoretic lower bound with the Bayesian Cram\'{e}r-Rao lower bound we derive for the case when the Bayes risk is the mean squared error. We illustrate the utility of the bounds through simulations by comparing them with the error performance of an expectation-maximization based inference algorithm proposed for the Bayesian BTL model.  We draw parallels between pairwise comparisons in the BTL model and inter-player games represented as edges in an Erd\H{o}s-R\'{e}nyi graph and analyze the effect of various graph structures on the lower bounds. We also extend the information-theoretic and Bayesian Cram\'{e}r-Rao lower bounds to the more general Bayesian BTL model which takes into account home-field advantage.
\end{abstract}
\section{Introduction}
Ranking systems are ubiquitous in daily life as they form integral parts of several applications, including electoral preference learning, personalized ad targeting, recommender systems, etc. A ranking system collates the opinions of its survey participants and obtains the true underlying ranking order that best agrees with the majority opinion, assuming that it exists. The ranking order corresponding to the majority opinion is often referred to as the \emph{consensus ranking}. 

When queried about the ranking order of $q$ items, the survey participants will usually share a list of $\ell\le q$ items in the order of preference. A large body of works consider \emph{permutations} of the set $\{1, \ldots, q\}$ as observed ranking orders, i.e., $\ell = q$,  and define a parameterized probability distribution function over the $q!$ permutations \cite{flignerdistance, meilaconsensus, MardenBook}. Several other works assume observations consisting of the top-$\ell$ rated items, where $\ell< q$, and derive inference algorithms for such parametric and non-parametric ranking models \cite{PlackettOrig,LuceOrig,MeilaBao}. Often the survey participants prefer providing quick responses in the form of pairwise preferences, especially if $q$ is large. Typically, such pairwise preferences are in response to queries of the form, \emph{``Is item $i$ better than item $j$?''}. These observations naturally arise in applications such as sports where two teams play against each other, elections where two candidates face-off, or social choice~\cite{EBApaper} etc. 

Amongst the ranking models for pairwise preferences \cite{MardenBook}, the Bradley-Terry-Luce (BTL) model is a popular, simple yet powerful model \cite{Zermelo1929, Ford1957, BTLOrig}. The BTL model associates a skill parameter to each item that is being compared. Several authors have addressed the problem of rank aggregation in the BTL model. In \cite{hunterMM}, the author uses the  minorization-maximization (MM) approach to infer the skill parameters of the BTL model. The rank centrality algorithm proposed in \cite{rankcentralityNegahban} is another popular approach, where the authors derive, using the theory of Markov chains and random walks, finite sample error rates between the skill parameters of the BTL model and those estimated by the algorithm. Counting algorithms such as Copeland counting~\cite{NiharSimpleRobust} and the weighted counting algorithm~\cite{Wauthier} have been also proposed for rank aggregation in the BTL model. In~\cite{ShivaniRajkumar}, the authors consider ranking under the BTL model along with several other models and obtain upper bounds on the sample complexity. The conditions for recovering the entries of the pairwise comparison matrix of a more general class of models, which is based on a strong stochastic transitivity property and includes the BTL model as a particular case, have also been derived in \cite{ShahICML2016}. 

As an alternative approach, by incorporating prior information into the  comparison model, Bayesian methods have also been applied for estimating the parameters of the BTL model. In fact, this approach has a long history in modeling animal behavior using the theory of dominance hierarchies~\cite{Drews}. In the case of animal behavior, maximum likelihood estimates of the skill parameters under the BTL model often do not converge to finite values (i.e., they are ill-conditioned), and the Bayesian methods are used as regularization techniques resulting in convergent (and well-conditioned) inference algorithms~\cite{Adams,Davidson,Leonard77}. More recent works have also investigated Bayesian preference learning in the setting where the pairwise comparisons are assumed to follow the probit model. This is a model in which each item is associated with a parameterized utility model based on a Gaussian process. For inference, gradient descent algorithms \cite{ChuGhahramani,MultitaskBirlutiu} and expectation propagation algorithms have been proposed \cite{SeegerScaCollabBPL}. 

A generalized Bayesian BTL model was introduced in \cite{CaronDoucet}. Here, the authors assign a Gamma distribution as a prior for the skill parameters. They show that by using a set of appropriate latent variables, it is possible to re-interpret the MM algorithms proposed by \cite{hunterMM}  as special instances of expectation-maximization (EM) algorithms. They propose such EM algorithms to infer the skill parameters in the basic BTL model and in several extensions such as the BTL model with home-field advantage and with ties. Here, we focus on this line of models.

\subsection{Main Contributions}
In this work, we derive lower bounds on the Bayes risks of estimators in the Bayesian BTL model described in \cite{CaronDoucet}, which also serve as lower bounds on their minimax risks.  More specifically, we use two separate lines of analyses  in Section \ref{sec:Main-Results}, and we obtain the following main results:
\begin{itemize}
\item In Section \ref{subsec:Main-Results-Minimax},  Theorem \ref{thm:Bayes-Risk-LB}  states a  family of information-theoretic lower bounds on the Bayes risks of estimators for norm-based distortion functions. For an $r$-norm to power $r$ distortion function, the theorem reveals that the Bayes risk dominates the function $n^{-r/2}$ asymptotically. The bounds given in  \eqref{eq:Bayes-Risk-LB-general} are obtained via the evaluation of a family of information-theoretic lower bounds proposed by Xu and Raginsky~\cite{Raginskypaper} and which we re-state in Theorem~\ref{thm:Raginsky}. The key step in our evaluation is the derivation of Proposition~\ref{prop:MI} to upper bound information-theoretic quantities associated to the model variables. 
\item In Section \ref{subsec:Main-Results-Bayesian-CRB}, Theorem \ref{thm:BCRB} provides the Bayesian Cram\'{e}r-Rao lower bound (BCRB) on the mean squared error (MSE) performance of estimators. 
\end{itemize}
After we present the lower bounds, we first discuss the effects of the hyper-parameters of the Gamma distributed prior on the lower bounds for two extreme cases of the parameter values. Then, to assess the tightness of the derived lower bounds, we illustrate their performance compared to the performance of the EM algorithm in \cite{CaronDoucet}. These discussions are presented in Section \ref{subsec:Main-Results-Discussions}. We note that \cite{ShahMinimax} has analyzed the estimation performance of inference algorithms in the BTL model. In contrast, we provide insights into the estimation performance in the Bayesian BTL models. 

As an application, we represent the pairwise comparison model using an Erd\H{o}s-R\'{e}nyi (ER) graph. In this representation, the comparison of a pair of items is viewed as a game between two players which induces an edge in the random graph. We analyze the lower bounds of Theorems \ref{thm:Bayes-Risk-LB} and \ref{thm:BCRB} to uncover the effect of graph structure on the bounds. In particular, given a fixed budget for the total number of comparisons, we answer the following questions in Section \ref{sec:Random-Graph}:
\begin{itemize}
\item[(q.1)] In a connected graph, how should one distribute edges in the graph, i.e., allocate the comparisons to pairs of items, such that the lower bounds are minimized. 
\item[(q.2)] Amongst all tree graphs (so the total number of edges is fixed and the graph is connected),  which tree structures minimizes and maximizes the lower bounds? 
\end{itemize}
The following answer to (q.1) is found in Section \ref{subsec:optimal-allocation-IT-Bounds} via Corollary \ref{cor:Minimax-optimal-graph}: All connected regular graph topologies minimize the information-theoretic lower bounds of Theorem \ref{thm:Bayes-Risk-LB}. In answering (q.2), we consider the two extremal tree graphs, namely the star graph with spokes emanating from a single node and the single-link chain graph. In Section \ref{subsec:optimal-allocation-IT-Bounds}, we further prove in Corollary \ref{cor:Minimax-optimal-tree} that, amongst all tree graphs, the star graph and the chain graph structures maximizes and minimizes, respectively, the the information-theoretic lower bounds of Theorems \ref{thm:Bayes-Risk-LB}.  Thus, we conclude that the chain graph structure of scheduling games leads to lower MSE. We also conjecture via basic simulations (for various values of $n$ and $k$) that the same conclusions hold for the BCRB of Theorem \ref{thm:BCRB}.  As a last point,  we briefly investigate whether the lower bounds we derived demonstrate phase transitions in the ER graph model.
 
Finally, we consider in Section \ref{sec:Home-Advantage} an extension of the basic Bayesian BTL model modified to account for home-field advantage in pairwise comparisons. For this model, also studied in \cite{CaronDoucet}, we carry similar lower bound derivations based on the same two techniques and state the results in Theorems \ref{thm:Bayes-Risk-LB-HA} and \ref{thm:Bayes-Risk-BCRB-HA}. Performance plots and conclusions drawn from the analyses are also provided.   

We defer most proofs to the Appendices or the supplementary material~\cite{supp}. 

\section{Preliminaries}\label{sec:preliminaries}
We first introduce some basic notations.
We define $[k]:=\{1,\ldots, k\}$.  Let $\mathcal{I}[k]:=\{(i, j): i, j\in[k], j\neq i\}$ denote the set of distinct item pairs and $\mathcal{I}_o[k]:=\{(i, j): i, j\in[k], i<j\}$ denote the ordered set of item pairs from the set $[k]$. We denote by $\mathds{1}\{\cdot\}$ the indicator function of a set. The superscript $T$ is used to indicate the matrix transpose operation. The $(i,j)^{\mathrm{th}}$ element of a matrix $\mathbf{M}$ is denoted as $[\mathbf{M}]_{i j}$ or $M_{i  j}$. The notations $\mathbb{R}$, $\mathbb{R}_{+}$, $\mathbb{R}_{++}$. and $\mathbb{N}$ are used as usual to indicate reals, non-negative reals, positive reals, and natural numbers, respectively.  The notation $\sim$ is used to mean ``distributed as'' and $\Expt[\cdot]$ denotes the expectation operator.  We will frequently come across two probability distributions. These are the binomial distribution, given by
$\mathcal{B}(k; n, q) := {n \choose k} q^{k} (1-q)^{n-k}$, for $k\in \{0, \ldots, n\}$,  where $n\in\mathbb{N}$  and $q\in [0, 1]$, and the Gamma distribution,  given by
\begin{equation}
\mathcal{G}(x; \alpha, \beta) = \displaystyle\frac{\beta^\alpha}{\Gamma(\alpha)} x^{\alpha-1}e^{-\beta x},
\end{equation}
for $x,\alpha, \beta\in\mathbb{R}_{++}$. The parameters $\alpha$ and $\beta$ are, respectively,   the {\em shape} and {\em rate} parameters  and $\Gamma(\cdot)$ is  the Gamma function. We denote the diagamma function  by $\psi(\cdot)=\Gamma'(\cdot)/\Gamma(\cdot)$, and  the Beta function by $B(x, y)$, for $x, y \in\mathbb{R}_{++}$. We use $O(\cdot)$ denote the Big-O notation. We also use the notation $\lesssim_x$ to say that a function is asymptotically less than or equal to another, i.e, $f(x)\lesssim_x g(x)$ holds if and only if $\limsup_{x\to\infty} f(x)/g(x) \leq 1$. Similarly,  $\gtrsim_x$ is used to denote the asymptotic inequality in the reverse direction. 

\subsection{The Bayesian BTL model}\label{subsec:Pre-Bayesian-BTL}
We now proceed with the description of the basic model and its integration into a Bayesian framework.  
\subsubsection{Ranking from pairwise comparisons}
Consider a collection of $k\geq 2$ items indexed by $[k]$. The outcomes of $n\in\mathbb{N}$ pairwise comparisons between the items of this collection consists of a record of the form:
\begin{equation}\label{eq:record}
\{(i_1,j_1,\ell_1),...,(i_n,j_n,\ell_n)\} \in\left(\mathcal{I}_o[k] \times\{0,1\}\right)^n ,
\end{equation}
 where  $(i_m, j_m)\in\mathcal{I}_o[k]$,  for each $m\in[n]$, indicates the indices of the item pairs being compared at the $m$-th comparison, and $\ell_m := \mathds{1}\{i_m \hbox{ is preferred over } j_m\}$ is the corresponding preference label. For each pair of items  $(i, j)\in\mathcal{I}_o[k]$, the problem of ranking from pairwise comparisons postulates the existence of underlying pairwise preference probabilities  such that item $i$ is preferred over item $j$ with probability $P_{ij}\in[0, 1]$ and the opposite is true with probability $P_{ji} =  1-P_{ij}$. Moreover,  the pairwise comparisons between item pairs  are assumed to be independent.  The pairwise preference probabilities  collectively form an underlying \textit{pairwise preference matrix} $\mathbf{P}$, and the class of all such matrices is given by:
\begin{equation}
\mathcal{P}:= \left\{\mathbf{P}\in[0,1]^{k\times k}: \parbox[c]{1.74 in}{$P_{ji} = 1 - P_{ij}, \forall (i, j)\in\mathcal{I}_o[k]$,  \vspace{0.04 in} \\ $P_{ii}=0, \forall i\in[k]$}    \right\}.
\end{equation}
The goal of ranking is to recover an accurate estimate of  $\mathbf{P}\in\mathcal{P}$ with respect to a desired norm. We will be particularly interested in the squared $L^2$-norm.

\subsubsection{Definition of the BTL model}
Multiple classes of statistical models for ranking have been proposed in the literature by imposing additional conditions on the structure of the permissible matrices $\mathcal{P}$ \cite{Agarwal2016}. The BTL model associates to each item $i\in[k]$ a skill parameter $\lambda_i\in\mathbb{R}_{++}$ such that 
\begin{equation}
P_{ij} := \displaystyle\frac{\lambda_i}{\lambda_i+\lambda_j},
\end{equation}
for all $i,j\in\mathcal{I}[k]$.  In other words, the task of a ranking algorithm here is to recover an accurate estimate of $\boldsymbol{\lambda}:= (\lambda_1, \ldots, \lambda_k)$ in the BTL model  governed by the following subclass of distributions: 
\begin{align}
\mathcal{P}_{\mathrm{BTL}} := \left\{  \mathbf{P}\in \mathcal{P} : \parbox[c]{1.74 in}{$\exists\, \boldsymbol{\lambda}\in\mathbb{R}_{++}^k \hbox{ s.t. } P_{ij}=\frac{\lambda_i}{\lambda_i+\lambda_j}$,  \vspace{0.04 in} \\ $\forall (i, j)\in\mathcal{I}_o[k]$}    \right\}.
\end{align}

From the definition of the class $\mathcal{P}_{\mathrm{BTL}}$, it can be seen that the parameter vector $\boldsymbol{\lambda}$ induces a family of conditional probability distributions $\{p(\cdot|\boldsymbol{\lambda}):\boldsymbol{\lambda}\in\mathbb{R}_{++}^k\}$ on the observation space $\{\mathcal{I}_o[k] \times\{0,1\}\}^n $.  In describing the induced probability distributions, it is sufficient and convenient to extract from the record in \eqref{eq:record} two quantities for any pair of items $(i, j)\in\mathcal{I}[k]$: The first is the number of comparisons in which element $i$ is preferred over element $j$,  which is denoted by $w_{ij}$, and the second is the total number of comparisons between elements $i$ and $j$,  which is denoted by $n_{ij}$.   Note that we necessarily have $n_{ij} = w_{ij}+w_{ji}$, for any $(i, j)\in\mathcal{I}[k]$,  and the total number of pairwise comparisons $n\in\mathbb{N}$ is given by 
\begin{equation}\label{eq:fixed-budget}
n= \displaystyle\sum_{(i,j)\in\mathcal{I}_o[k]}n_{ij} = \frac{1}{2}\displaystyle\sum_{(i,j)\in\mathcal{I}[k]}n_{ij}.
\end{equation}
In the scope of this work, we will further assume that $\mathbf{N} := (n_{ij})\in\mathbb{N}^{k\times k}$ is a matrix that is fixed {\em a priori}, and the comparisons are performed accordingly.\footnote{The question of how to ``optimally'' choose $\mathbf{N}$ for a fixed budget $n$ will be addressed later in Section \ref{sec:Random-Graph} in the context of random graphs.}
Now, a data sample can be described by the matrix $\mathbf{W} := (w_{ij})\in\mathbb{N}^{k\times k}$.
Correspondingly, we let $\boldsymbol{\Omega} = (\Omega_{ij})\in\mathbb{N}^{k\times k}$ denote the random data matrix, i.e., $w_{ij}$ is assumed to be the realization of a random variable $\Omega_{ij}$, for all $(i, j)\in\mathcal{I}[k]$.
 Then, one can show that, for each $\boldsymbol{\lambda}\in\mathbb{R}_{++}^k$, the basic BTL model assumption results in the following conditional distributions:
 \begin{equation}\label{eq:prob-D-given-lambda}
\boldsymbol{\Omega} |\boldsymbol{\lambda}\sim  p(\mathbf{W}|\boldsymbol{\lambda})   
  =  \prod_{(i, j)\in\mathcal{I}_o[k]}  \mathcal{B}(w_{ij}; n_{ij}, P_{ij}),
 \end{equation}
where $\Omega_{ij}|\lambda_i, \lambda_j \sim p(w_{ij}|\lambda_i, \lambda_j)  = \mathcal{B}(w_{ij}; n_{ij}, P_{ij})$.  See Lemma 1 in supplementary material~\cite{supp} for a proof of \eqref{eq:prob-D-given-lambda}.
 
\subsubsection{Bayesian estimation framework}
In the Bayesian estimation framework, the unknown parameter vector is treated as a random vector  $\boldsymbol{\Lambda} := (\Lambda_1, \ldots, \Lambda_k) \in\mathbb{R}_{++}$ and the parameter space is endowed with a prior distribution $p(\boldsymbol{\lambda})$ on $\boldsymbol{\Lambda}$. Then, it is assumed that, for a given realization $\boldsymbol{\Lambda} = \boldsymbol{\lambda}$ and for a fixed $\mathbf{N}$,  a data sample $\mathbf{W}$ is generated according to the probability distribution $p(\mathbf{W}|\boldsymbol{\lambda})$. The joint distribution of the pair $(\boldsymbol{\Omega}, \boldsymbol{\Lambda})$ for fixed $\mathbf{N}$ is now uniquely determined by $p(\boldsymbol{\lambda}, \mathbf{W}) = p(\boldsymbol{\lambda}) p(\mathbf{W}|\boldsymbol{\lambda})$. 
In this framework,  the {\em Bayes risk} for estimating $\boldsymbol{\Lambda}$ from $\boldsymbol{\Omega}$ with respect to a given distortion function $d: \mathbb{R}_{++}^k \times \mathbb{R}_{++}^k \to \mathbb{R}^+$ is defined as 
\begin{equation}\label{eq:Bayes-Risk}
R_{\mathrm{B}} :=\displaystyle\inf_{\boldsymbol{\varphi}} \Expt[d(\boldsymbol{\Lambda}, \boldsymbol{\varphi}(\boldsymbol{\Omega}))],
\end{equation}
where $\boldsymbol{\varphi}(\cdot) : \mathbb{N}_{++}^k \times \mathbb{N}_{++}^k \to \mathbb{R}_{++}^k $ is an estimator of $\boldsymbol{\Lambda}$.

\subsubsection{Choice of  prior distributions}
The works \cite{CaronDoucet, guiver2009, gormley_2009b}, which perform Bayesian estimation for the basic BTL model or its generalizations, assign a Gamma distributed prior  $\Lambda_i\sim p(\lambda_i) = \mathcal{G}(\lambda_i; a_i, b_i)$ to each skill parameter $i\in[k]$, where $\mathbf{a}:=(a_i), \mathbf{b}:= (b_i)\in\mathbb{R}_{++}^k$.\footnote{Prior works take $a_i= a$, $b_i=b$, for all $i\in[k]$, but we introduced the more general version as some of our results are also applicable to this case.} We will be assuming these priors throughout this paper. So, we let
\begin{equation}\label{eq:gamma-prior}
\boldsymbol{\Lambda}\sim p(\boldsymbol{\lambda}) = \displaystyle\prod_{i\in[k]} p(\lambda_i)=\displaystyle\prod_{i\in[k]}\mathcal{G}(\lambda_i; a_i, b_i),
\end{equation}
and by \eqref{eq:prob-D-given-lambda} and \eqref{eq:gamma-prior}, we get the following expression:
\begin{equation}\label{eq:prob-lambda-D}
p(\boldsymbol{\lambda}, \mathbf{W})  = \prod_{(i,j)\in\mathcal{I}_o[k]}  \mathcal{B}(w_{ij}; n_{ij}, P_{ij}) \prod_{i\in[k]} \mathcal{G}(\lambda_i; a_i,b_i).
\end{equation}

\subsubsection{Introducing Latent Random Variables}
The  assumption in \eqref{eq:gamma-prior} turns out to be a convenient choice, justified by what is called in the literature ``the Thurstonian interpretation'' of the BTL model \cite{Diaconis88}. In fact,  the probability that  an item is preferred over another one in a pairwise comparison in the BTL model can be naturally seen as being determined by the shortest of two exponentially distributed arrival times with rate parameters given by the respective skill parameters of the items. Namely, the correspondence $P_{ij} = \Prob[\Upsilon_{si} < \Upsilon_{sj}]$ can be established, for each pair $(i, j)\in\mathcal{I}_o[k]$ and for all $s=1,\ldots n_{ij}$, by defining the latent random variables $\Upsilon_{si}\sim\mathcal{E}(\lambda_i)$ and $\Upsilon_{sj}\sim\mathcal{E}(\lambda_j)$, where $\mathcal{E}(\lambda)$ is the exponential distribution with rate $\lambda$.\footnote{It should be clear that from the realizations of the random arrival times, one can obtain the data sample $\mathbf{W}$.}
 
For getting faster rates of convergence for the EM and the data augmentation algorithms they propose for performing Bayesian inference, Caron and Doucet \cite{CaronDoucet} introduced the following set of latent random variables:
\begin{equation}\label{eq:Zij}
Z_{ij} = Z_{ji} : = \displaystyle\sum_{s=1}^{n_{ij}} \min\{\Upsilon_{si}, \Upsilon_{sj}\},
\end{equation}
for $(i, j)\in\mathcal{I}_o[k]$. This new set of latent variables will be useful in our information-theoretic lower bound derivations. We define the random matrix $\mathbf{Z} := (Z_{ij}) \in\mathbb{R}^{k\times k}$ and denote its realization by $\boldsymbol{\zeta} := (\zeta_{ij}) \in \mathbb{R}^{k\times k}$. From \cite[Eq.~(2.1)]{CaronDoucet}, 
\begin{equation}
Z_{ij}|\lambda_i, \lambda_j \sim p(\zeta_{ij}|\lambda_i, \lambda_j)= \mathcal{G}(\zeta_{ij}; n_{ij}, \lambda_i+\lambda_j), \label{eq:prob-Z-given-D-lambda}
\end{equation}
  for all $(i, j)\in\mathcal{I}[k]$.

\subsection{Lower Bounds on the Bayes Risk}\label{subsec:Preliminaries-BayesRisk}
Next in line is the presentation of the tools we use to compute lower bounds on the Bayes risk of estimators.  Note that our lower bounds on the Bayes risk automatically serve as lower bounds on the minimax risk---a more general notion of risk associated to estimation problems given in our context by 
 \begin{equation}
 R_{\mathrm{M}} := \displaystyle\inf_{\boldsymbol{\varphi}}  \displaystyle\sup_{\boldsymbol{\Lambda}\sim p(\boldsymbol{\lambda})} \Expt[d(\boldsymbol{\Lambda}, \boldsymbol{\varphi}(\boldsymbol{\Omega}))].
 \end{equation}
 Since the minimax risk is computed by choosing an estimator that minimizes the maximum of the Bayes risk defined in \eqref{eq:Bayes-Risk}, $R_{\mathrm{M}} \geq R_{\mathrm{B}}$ always holds. Although several techniques exist to compute lower bounds on the minimax risk of estimation and optimization problems (see for instance \cite{Yu1997}), our focus will be on computing lower bounds on the Bayes risk.
 
\subsubsection{Information-theoretic lower bounds}
The lower bounds we derive in Sections~\ref{subsec:Main-Results-Minimax} and~\ref{subsec:Home-Advantage-Minimax} will make use of the following result from \cite{Raginskypaper} involving information-theoretic quantities.
\begin{thm}\cite[Theorem 3]{Raginskypaper}\label{thm:Raginsky} 
Let $\lVert\cdot\rVert$ be an arbitrary norm in $\mathbb{R}^k$ and let $r \geq 1$. The Bayes risk for estimating the parameter $\mathbf{X}\in\mathbb{R}^k$ based on the sample $\mathbf{Y}$ with respect to the distortion function $d(x,\hat{x})=\lVert x-\hat{x}\rVert^r$ satisfies
\begin{multline}\label{eq:Raginsky-bound}
R_{\mathrm{B}} \geq \displaystyle\sup_{p(\mathbf{T}|\mathbf{X}, \mathbf{Y})} \displaystyle \frac{k}{re}\left(V_k \Gamma\left(1+\displaystyle\frac{k}{r}\right)\right)^{-r/k}  \\
\times e^{-\left(I(\mathbf{X};\mathbf{Y}|\mathbf{T})- h(\mathbf{X}|\mathbf{T})\right)r/k},
\end{multline} 
where $V_k$ denotes the volume of the unit ball in $(\mathbb{R}^k, \lVert\cdot\rVert)$. $I$ and $h$ denote the (conditional) mutual information and (conditional) entropy, respectively.
\end{thm}

\subsubsection{Cram\'{e}r-Rao type bounds on the Bayes risk}\label{subsec:Preliminaries-BCRB}
Consider a general estimation problem where the unknown vector $\mathbf{X} \in \mathbb{R}^k$ can be split into sub-vectors $\mathbf{X} =
[\mathbf{X}_{\mathrm{\textbf{r}}}^T,~\mathbf{X}_{\mathrm{\textbf{d}}}^T]^T$, where $\mathbf{X}_{\textbf{r}} \in \mathbb{R}^m$  consists of \emph{random} parameters distributed according to a known distribution, and $\mathbf{X}_{\textbf{d}}\in \mathbb{R}^{k-m}$ consists of \emph{deterministic} parameters. Let $\bm{\varphi}(\mathbf{Y})$ denote an estimator of $\mathbf{X}$ as a function of the observations $\mathbf{Y}$. Recall that the MSE matrix is defined as $\mathbf{E}^{\mathbf{X}} := \mathbb{E}\left[(\mathbf{X} - \bm{\varphi}(\mathbf{Y}))(\mathbf{X} - \bm{\varphi}(\mathbf{Y}))^T\right]$. 
The first step in obtaining Cram\'{e}r-Rao-type lower bounds \cite{Kay} is to derive the Fisher Information Matrix (FIM). In this paper, we use the notation  $\mathbf{I}^{\mathbf{X}}$ to represent the FIM under the different modeling assumptions. Typically, $\mathbf{I}^{\mathbf{X}}$ is expressed in terms of the individual blocks of submatrices, where the $(i,j)^{\mathrm{th}}$ block is given by 
\begin{equation}\label{Infomat_def_gen}
[\mathbf{I}^{\mathbf{X}}]_{ij} := -\mathbb{E} \left[\left(\nabla_{\mathbf{X}}\right)_i\left(\nabla_{\mathbf{X}}\right)_j^T \log \pr(\mathbf{Y},\mathbf{X}_{\mathrm{\textbf{r}}}|\mathbf{X}_{\mathrm{\textbf{d}}}) \right],
\end{equation}
where $\nabla_{\mathbf{X}}$ denotes the gradient with respect to the vector  $\mathbf{X}$. Then, assuming that the MSE matrix $\mathbf{E}^{\mathbf{X}}$ exists and the FIM $\mathbf{I}^{\mathbf{X}}$ is non-singular, a lower bound on $\mathbf{E}^{\mathbf{X}}$ is given by
\begin{equation}
\mathbf{E}^{\mathbf{X}} \succeq \left(\mathbf{I}^{\mathbf{X}}\right)^{-1}.
\end{equation}
For example, when $\mathbf{X}_{\mathrm{\textbf{r}}} \neq \emptyset$ and $\mathbf{X}_{\mathrm{\textbf{d}}} = \emptyset$, $\mathbf{I}^{\mathbf{X}}$ represents the Bayesian Information matrix (BIM) and the corresponding lower bound on the MSE matrix is called the BCRB. When $\mathbf{X}_{\mathrm{r}} \neq \emptyset$ and $\mathbf{X}_{\mathrm{d}} \neq \emptyset$, $\mathbf{I}^{\mathbf{X}}$ represents the Hybrid Information Matrix (HIM), and the corresponding lower bound on the MSE matrix is called as the hybrid Cram\'{e}r-Rao bound (HCRB). Finally, when the squared $L^2$ norm is used as the distortion measure, the Bayes risk  can be lower bounded by the trace of the inverse of the FIM. 

\section{Main Analytical Results}\label{sec:Main-Results}
In this section, we present our main results following from the information-theoretic and Cram\'{e}r-Rao analyses. 
\subsection{Information-Theoretic Lower Bounds}\label{subsec:Main-Results-Minimax}
The next theorem states the main result of this subsection. Its proof will be given at the end.
\begin{thm}\label{thm:Bayes-Risk-LB}
Consider the Bayesian BTL model introduced in Section \ref{subsec:Pre-Bayesian-BTL}. Let $\lVert \cdot \rVert$ denote an arbitrary norm in $\mathbb{R}^k$. For any  $r\geq 1$, let  $d(\boldsymbol{\lambda}, \boldsymbol{\widehat{\lambda}}) = \lVert \boldsymbol{\lambda} - \boldsymbol{\widehat{\lambda}}\rVert^r$ be the distortion function, where $\boldsymbol{\widehat{\lambda}}:= \boldsymbol{\varphi}(\mathbf{W})$ is an estimator of $\boldsymbol{\lambda}$ based on data sample $\mathbf{W}$ for a fixed $\mathbf{N}$.   For all $i\in[k]$, let
\begin{equation}\label{eq:ni}
n_i:= \frac{1}{2}\sum_{j\in[k]\setminus\{i\}}n_{ij}.
\end{equation}
Then, the Bayes risk $R_{\mathrm{B}}$ defined  in \eqref{eq:Bayes-Risk} for estimating $\boldsymbol{\lambda}\in\mathbb{R}_{++}^k$ is asymptotically lower bounded by\footnote{ The notation $\gtrsim_{n_i}$ means that the LHS asymptotically dominates the RHS as $n_i\to\infty$ {\em for all} $i\in[k]$.}
\begin{equation}
\label{eq:Bayes-Risk-LB-general} R_{\mathrm{B}}  \hspace{1mm}\gtrsim_{n_i}\frac{k}{re}\left(V_k\Gamma\left(1+\frac{k}{r}\right)\right)^{-r/k} e^{-rE_{\mathrm{BTL}}(\mathbf{N}, \mathbf{a}, \mathbf{b}) },
\end{equation}
where $V_k$ is the volume of the unit ball in $(\mathbb{R}^k, \lVert\cdot\rVert)$, 
\begin{multline}\label{eq:E1}
E_{\mathrm{BTL}}(\mathbf{N}, \mathbf{a}, \mathbf{b}) 
:= \frac{1}{k}\sum_{i\in [k]} \Bigg( - \frac{1}{2}\log{(2\pi)} + \log{b_i}-\psi(a_i) \\
+\frac{1}{2}\log{ \left(a_i + n_i\right)}\Bigg).
\end{multline}
\vspace{-5mm}
\end{thm}
\begin{cor}\label{cor:Bayes-Risk-LB}
If $a_i=a$ and $b_i = b$, for each $i\in[k]$, one can further lower bound the expression in \eqref{eq:Bayes-Risk-LB-general} via Jensen's inequality.
%\begin{multline}
%R_{\mathrm{B}} \label{eq:Bayes-Risk-LB-2}\gtrsim_n \frac{k}{re}\left(V_k\Gamma\left(1+\frac{k}{r}\right)\right)^{-r/k} \times\\
% \frac{\exp\big\{-r \left( - \frac{1}{2}\log{(2\pi)} + \log{b}-\psi(a)\right)\big\}}{\sqrt[r/2]{a/k +n}}.
%\end{multline}
Consequently, for the $L^1$ norm ($r=1$),  we get:
\begin{equation}\label{eq:Bayes-Risk-LB-L1}
R_B \gtrsim_n  \sqrt{\frac{\pi}{2}} e^{-(\log{b}-\psi(a)+1)} \displaystyle\frac{k}{\sqrt{a/k +n}},
\end{equation}
and for the squared $L^2$ norm ($r=2$), we get
\begin{equation}\label{eq:Bayes-Risk-LB-L2}
R_B \gtrsim_n e^{-2(\log{b}-\psi(a))-1} \displaystyle\frac{k}{a/k +n}.
\end{equation} 
\end{cor}
In proving Theorem \ref{thm:Bayes-Risk-LB}, we will use Theorem \ref{thm:Raginsky} and the result we introduce in the next proposition.
\begin{prop}\label{prop:MI}
For the Bayesian BTL model introduced in Section \ref{subsec:Pre-Bayesian-BTL}, we have
\begin{equation}\label{eq:MI-Bound}
\frac{1}{k}\left(I(\boldsymbol{\Lambda}; \boldsymbol{\Omega}\mathbf{Z}) - h(\boldsymbol{\Lambda})\right)   \lesssim_{n_i}  E_{\mathrm{BTL}}(\mathbf{N}, \mathbf{a}, \mathbf{b}),
\end{equation}
where $n_i$ is defined in \eqref{eq:ni} and $E_{\mathrm{BTL}}(\mathbf{N}, \mathbf{a}, \mathbf{b})$ in \eqref{eq:E1}.  
\end{prop}
The proof of Proposition \ref{prop:MI} is given in Appendix \ref{app:proofProp}. Now, we are ready to prove the theorem.
\begin{IEEEproof}[Proof of Theorem \ref{thm:Bayes-Risk-LB}]
We first observe that 
\begin{equation}
R_{\mathrm{B}} \geq \inf_{\boldsymbol{\varphi}'} \Expt[\ell(\boldsymbol{\Lambda}, \boldsymbol{\varphi}'(\boldsymbol{\Omega}, \mathbf{Z}))]=:R_{\mathrm{B}}' .
\end{equation}
Now,   taking $\mathbf{X} \leftarrow \boldsymbol{\Lambda}$, and $\mathbf{Y} \leftarrow (\boldsymbol{\Omega}, \mathbf{Z})$ in Theorem \ref{thm:Raginsky}, the proof of the claimed asymptotic lower bound in Theorem~\ref{thm:Bayes-Risk-LB} follows by lower bounding $R_{\mathrm{B}}'$ via the unconditional version of the lower bound in \eqref{eq:Raginsky-bound} and then using the relation \eqref{eq:MI-Bound} derived in Proposition \ref{prop:MI}.
\end{IEEEproof}

\subsection{Bayesian Cram\'{e}r-Rao Lower Bound}\label{subsec:Main-Results-Bayesian-CRB}
In the next theorem, we state the BCRB, which is a well-known lower bound on the MSE of an estimator. In contrast to the family of information-theoretic lower bounds derived in the previous section, the BCRB does not require the auxiliary variable $\mathbf{Z}$. The proof of the theorem is given in Appendix \ref{app:prf_bcrb}.

\begin{thm}\label{thm:BCRB}
For the Bayesian BTL model introduced in Section \ref{subsec:Pre-Bayesian-BTL}, the entries of the  BIM are given by
\begin{align}\label{eq:BIM-ii}
&[\matI^{\bmLambda}]_{i,i} = (a_i-1) T_1(a_i,b)  +   \sum_{j\in[k]\setminus \{i\}}  n_{ij} T_2(a_i,a_j,b),
\end{align}
\begin{equation}
 [\matI^{\bmLambda}]_{i,j}=  - n_{ij} T_3(a_i,a_j,b), 
\end{equation}
for $i\in[k]$ and  for $(i, j)\in\mathcal{I}[k]$, where 
\begin{align}\label{eq:T1}
T_1(a_i,b) & := \Ex\left[\Lambda_i^{-2}\right] = \frac{b^2\Gamma(a_i-2)}{\Gamma(a_i)},\\
T_2(a_i,a_j,b)& :=  \frac{b^2 (a_i-2)\Gamma(a_i-2)}{\Gamma(a_i)} \nonumber\\*
\label{eq:T2}&\!\!\!\!\times\left[\frac{a_j}{ a_i+a_j - 2 } - \frac{\Gamma(a_j+1)}{(a_i+a_j - 1)\Gamma(a_j)}\right],\\
T_3(a_i,a_j,b)&:= \frac{b^2(a_i - 1)\Gamma(a_i-1)}{\Gamma(a_i)} \nonumber\\
\label{eq:T3}&\times\left[\frac{(a_j - 1)\Gamma(a_j-1)}{\Gamma(a_j)(a_i+a_j-1)} - \frac{1}{ a_i+a_j-2 }\right].
\end{align}
The BCRB  on the MSE matrix $\mathbf{E}^{\bmLambda}$ of the unknown random skill parameter vector $\bmLambda$ is given by $\mathbf{E}^{\bmLambda} \succeq (\mathbf{I}^{\bmLambda})^{-1}$, and the Bayes risk with squared $L^2$ norm is lower bounded as
\begin{equation}\label{eq:BayesRiskBCRB}
R_{\mathrm{B}} \geq \textnormal{Tr}((\mathbf{I}^{\bmLambda})^{-1}).
\end{equation}
 \end{thm}

\subsection{Discussions}\label{subsec:Main-Results-Discussions}
In a given statistical model, lower bounds on the Bayes risk of estimators help to characterize their fundamental performance limits. Any specific algorithm we run cannot perform better than the algorithm-independent fundamental limit, and thus naturally, than any of its lower bounds.  
We next present some properties of the lower bounds we derived for the Bayesian BTL model.

\subsubsection{Effect of priors} To simplify the discussion, we let $a_i=a$ and $b_i = b$   for each $i\in[k]$. In \cite{CaronDoucet}, the prior~\eqref{eq:gamma-prior} is chosen such that $b=ak-1$, for $a\in\mathbb{R}_{++}$ and $k\in\mathbb{N}$. This choice ensures that $\sum_{i\in[k]}\lambda_i = 1$, and it is justified by the fact that $b$ acts as a scaling parameter with no influence on inference~\cite[Section 5]{CaronDoucet}. This latter observation is reflected as well in the lower bounds of Theorems \ref{thm:Bayes-Risk-LB} and \ref{thm:BCRB} which depend on $b$ only as a multiplicative scaling factor given by $1/b^2$. In fact, the BCRB given in Theorem \ref{thm:BCRB} can be expressed as a function of $a/b^2$, i.e., the variance of the prior distribution in \eqref{eq:gamma-prior}. Let us next examine the behavior of the derived lower bounds in two extreme cases of the mean over variance ratio of the Gamma prior   in \eqref{eq:gamma-prior}. As this ratio is given by $b$, we consider the cases $b\to 0^{+}$ and $b \to \infty$. We note that the family of information-theoretic lower bounds  in \eqref{eq:Bayes-Risk-LB-general}  and the BCRB in \eqref{eq:BayesRiskBCRB} both tend to infinity when $b \to 0^{+}$ and tend to $0$ when $b \to \infty$. 

\begin{figure}[t] 
\begin{center}
\includegraphics[scale=0.45]{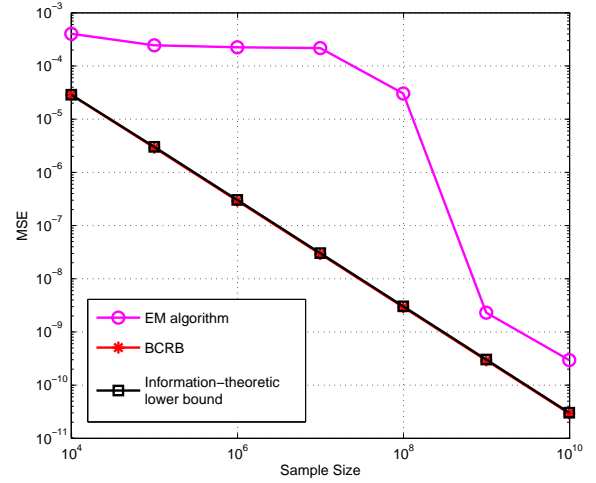}
\caption{MSE ($L^2$ error) performance of the EM algorithm and the information-theoretic and BCRB lower bounds of Theorems \ref{thm:Bayes-Risk-LB} and \ref{thm:BCRB}, respectively. Figure is generated for $k=100$ items. The parameters of the prior distribution in \eqref{eq:gamma-prior}  are chosen as $a=5$ and $b=ak-1$.}
\label{fig:MSE_EM_BCRB_Raginsky}
\end{center}
\squeezeup
\end{figure}

%\begin{figure}[t]
%\begin{center}
%\includegraphics[scale=0.45]{Figure2k100.eps}
%\caption{$L^1$ error performance of the EM algorithm and the  information-theoretic lower bound  of Theorem \ref{thm:Bayes-Risk-LB} .   Figure is generated for $k=100$ items. The parameters of the prior distribution in \eqref{eq:gamma-prior}  are chosen as $a=5$ and $b=ak-1$.}
%\label{fig:L1_EM_Raginsky}
%\end{center}
%\squeezeup
%\end{figure}
\subsubsection{Performance of Bounds} 
We now present some simulation results  to assess the tightness of our lower bounds. Fig.~\ref{fig:MSE_EM_BCRB_Raginsky} displays plots of the information-theoretic and BCRB lower bounds on the Bayes risk for the squared $L^2$ norm together with the MSE performance of the EM algorithm proposed by \cite{CaronDoucet} for $k=100$ items. Note that to simulate the MSE performance of the EM algorithm, we sampled the skill parameters as in \eqref{eq:gamma-prior} and the number of times an item is preferred over another one as in \eqref{eq:prob-D-given-lambda}, and we used the code provided by \cite{CaronDoucet} in their supplementary material. In general, we expect the information-theoretic lower bound to be smaller than the BCRB since the former has been derived by including the latent random matrix $\mathbf{Z}$ into the Bayesian estimation framework.\footnote{Any additional information regarding data can only decrease the lower bounds on the Bayes risk.} Nevertheless, we see from Fig.~\ref{fig:MSE_EM_BCRB_Raginsky} that the difference is negligible for $k=100$ items. In addition, we also see from the figure that the performance of the EM algorithm approaches the lower bounds as the number of samples increases. Thus, it appears that the bounds are increasingly tight as the sample size $n\to\infty$. We emphasize that this conclusion we draw experimentally holds regardless of the existence of global optimum guarantees for the EM algorithm of \cite{CaronDoucet}. In fact, our lower bounds are also valid for any instance of any algorithm, including those with a potentially lower MSE than the specific algorithm we run. 

Finally, we make some remarks concerning the finite sample performance of our lower bounds. We note that the BCRB of Theorem \ref{thm:BCRB} is already non-asymptotic. Regarding the family of information-theoretic lower bounds of Theorem \ref{thm:Bayes-Risk-LB}, we note that although they are asymptotic, this is only due to using Stirling's approximation in the derivations. In fact, Theorem \ref{thm:Bayes-Risk-LB} follows from Theorem \ref{thm:Raginsky}, which is non-asymptotic. The Stirling's approximation, which is known to be accurate even for small values of its argument, helped us to obtain a simple yet meaningful bound from which we can obtain more insights into the problem. In particular, as we will see next, it allows us to answer the questions posed in the Introduction.

\section{Effect of Graph Structure on Bounds}\label{sec:Random-Graph}

In any ranking procedure, the subset of the pairs of items  being compared induces  a comparison graph. Let $G := ([n],E)$ be a comparison graph such that if the item pair $(i,j)\in\mathcal{I}_0[k]$ belongs to the edge set $E$ with edge weight $n_{ij}\in\mathbb{N}$, then the items $i$ and $j$ are being compared $n_{ij}$ times. In this section, we investigate the effect of the graph structure on the lower bounds derived in the previous section. More specifically, we explore graph structures in the context of how to design experiments in pairwise comparisons in ranking to minimize the distortion, and we answer the questions (q.1) and (q.2) we posed in the Introduction.  The analysis can be used as a guideline in applications where the total number of pairwise comparisons $n$ is given, but the choice of the pairs to be compared has to be designed as part of the ranking procedure. 
 
 \subsection{Optimal Edge Allocations}\label{subsec:optimal-allocation-IT-Bounds}
 The next corollary identifies the optimal connected graph topologies arising from Theorem \ref{thm:Bayes-Risk-LB}.
\begin{cor}\label{cor:Minimax-optimal-graph}
Given a fixed budget for $n$, as defined in \eqref{eq:fixed-budget}, the minimum of the lower bounds on the Bayes risk in \eqref{eq:Bayes-Risk-LB-general} is achieved by the following water-filling solution for $n_i$ defined in \eqref{eq:ni}:
\begin{equation}
n_i = (\mu-a_i)^+,
\end{equation}
 for any $i\in[k]$, where $\mu$ is chosen so that
$\sum_{i\in[k]}  (\mu-a_i)^+ = n$.
\end{cor}
\begin{IEEEproof}
It is easy to see that the allocation of $n_i$'s, for all $i\in[k]$, which maximizes  $E_{\mathrm{BTL}}(\mathbf{N}, \mathbf{a}, \mathbf{b})$ defined in \eqref{eq:E1}, and thus minimizes the lower bounds in  \eqref{eq:Bayes-Risk-LB-general}, is given by the water-filling solution, since this optimization corresponds to the problem of maximizing $\sum_{i\in[k]}\frac{1}{2}\log{\left(a_i + n_i\right)}$, subject to the constraints $\sum_{i\in[k]} n_i = n$ and $n_i\in\mathbb{N}$, see for instance the discussion in \cite[Chapter 9.4]{Cover2006}. 
\end{IEEEproof}

Let us next consider the class of tree graphs, which are amongst the most simple graph topologies. Amongst all tree graphs with $k$ nodes and $k-1$ edges,  we focus on two extremal tree structures, the first one being the star graph which has one central node with edges to every other node, and the second one being the chain graph which consists of an arbitrary ordering of the $k$ nodes with edges only between pairs of neighbors. The next corollary identifies the extremal tree topologies arising from Theorem \ref{thm:Bayes-Risk-LB}. Its proof is provided in Appendix \ref{app:prf_tree}.  
\begin{cor}\label{cor:Minimax-optimal-tree}
Suppose that $a_i=a$, for all $i\in[k]$. Amongst all tree graphs with a fixed budget for $n$, as defined in \eqref{eq:fixed-budget}, the maximum and minimum values of the family of lower bounds on the Bayes risk in \eqref{eq:Bayes-Risk-LB-general}  are achieved by the extremal star and chain graphs, respectively. 
\end{cor}

\begin{figure}[t]
\begin{center}
\includegraphics[scale=0.45]{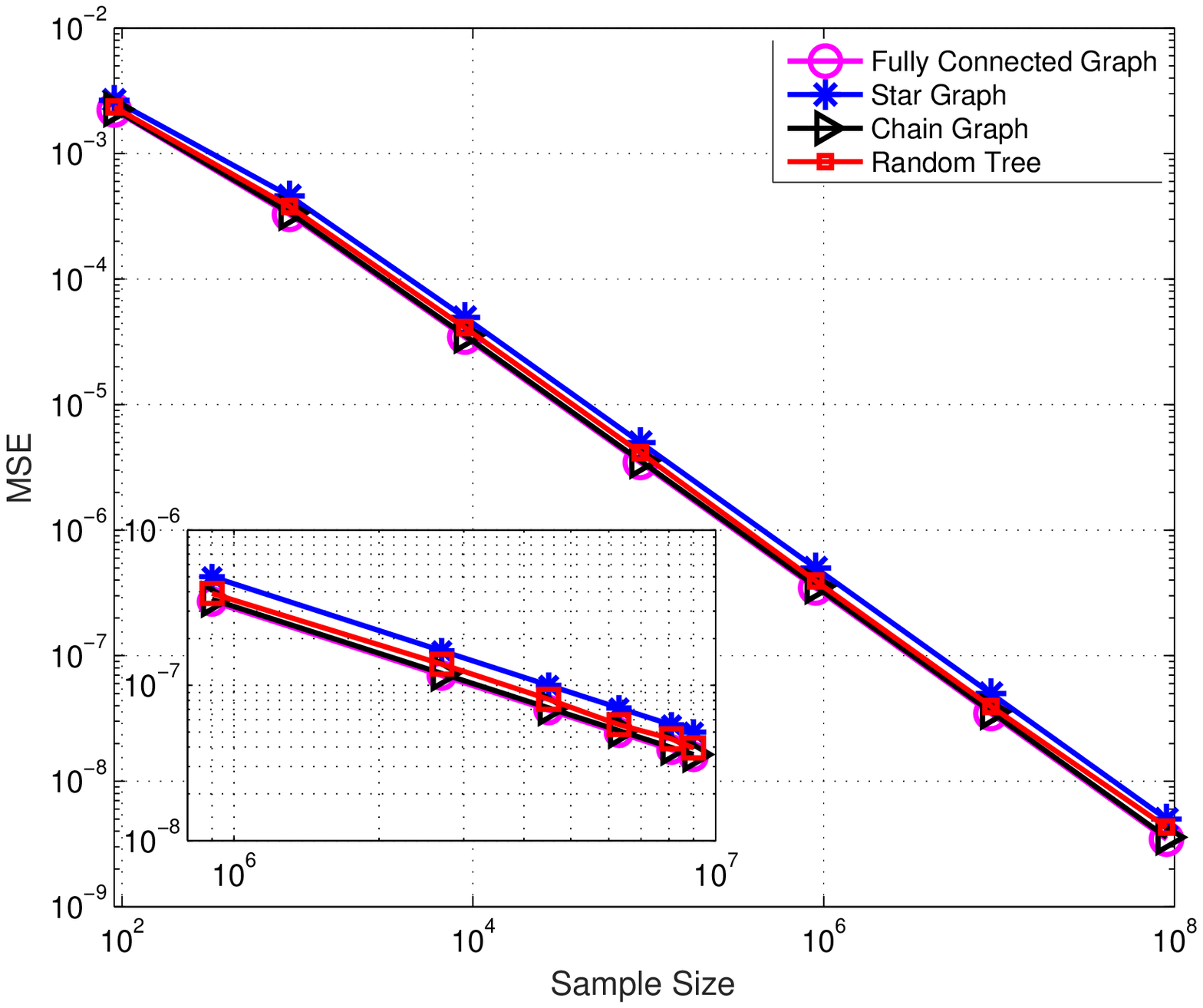}
\includegraphics[scale=0.45]{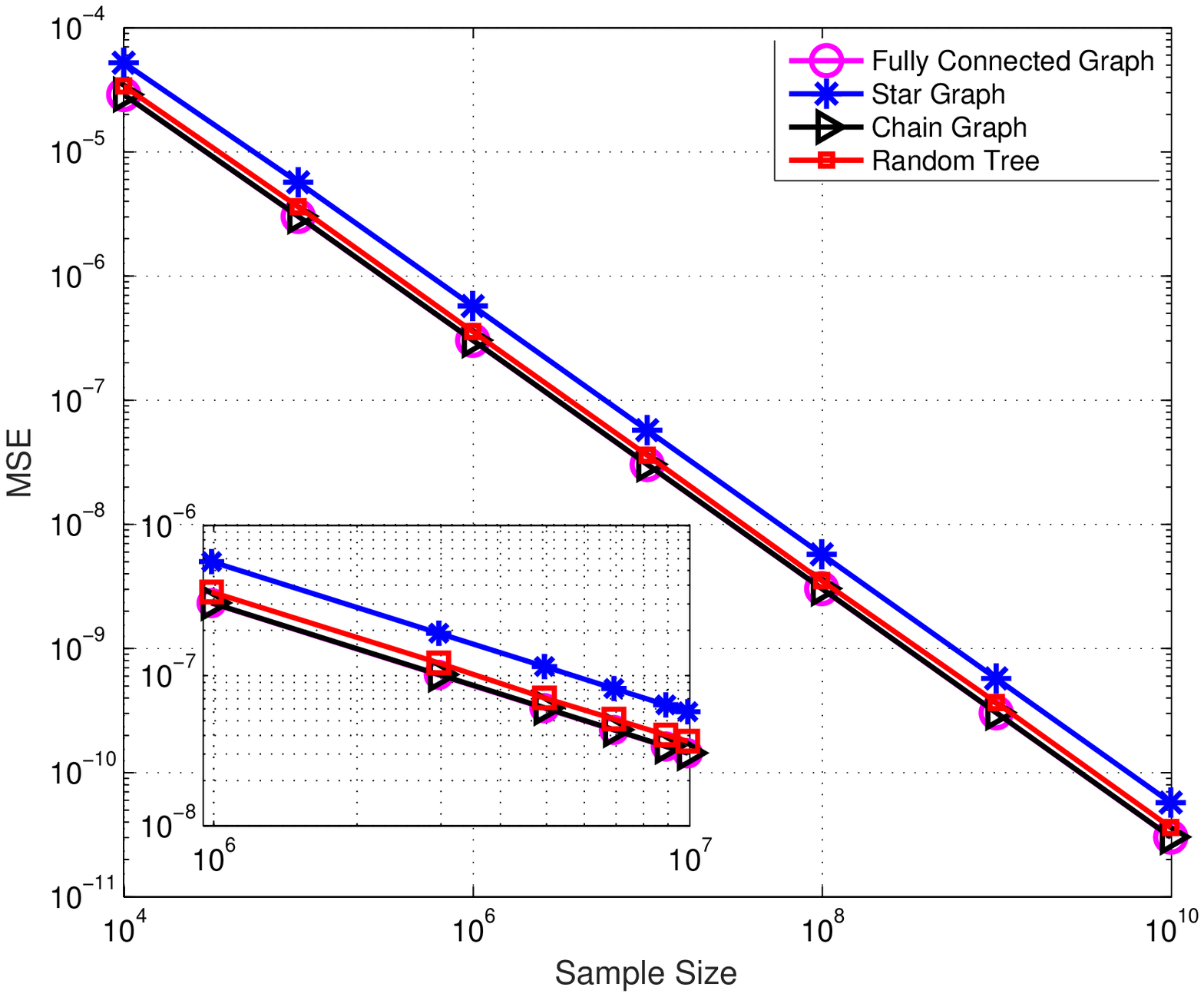}
\caption{Information-theoretic lower bounds of Theorem \ref{thm:Bayes-Risk-LB} for the squared $L^2$-norm as a function of the number of samples for different graph topologies. Top figure is generated for $k=10$ items, and bottom figure for $k=100$ items. The parameters of the prior distribution in \eqref{eq:gamma-prior}  are chosen as $a=5$ and $b=ak-1$.}
\label{MinimaxGraphs}
\end{center}
\squeezeup
\end{figure}

Based on the last two corollaries, we obtain the following answers to (q.1) and (q.2) we posed in the Introduction:
\begin{itemize}
\item[(a.1)] Given a fixed budget $n$, as defined in \eqref{eq:fixed-budget}, and $a_i=a$, for all $i\in[k]$, Corollary \ref{cor:Minimax-optimal-graph} implies that, amongst all connected graphs, any connected \emph{regular} graph results in an optimal allocation minimizing the lower bounds on the Bayes risk in  \eqref{eq:Bayes-Risk-LB-general}. One such graph is the fully connected graph with an equal number of pairwise comparisons with $n_i = n/k$ per node, for all $i \in[k]$, and $n_{ij} = 2n/\left(k(k-1)\right)$ per edge, for all $(i, j)\in\mathcal{I}[k]$. Another one is the cycle graph with  an equal number of pairwise comparisons $n_{i(i+1)} = n_{1k} = n/k$ per edge, for all $i\in[k-1]$. 
\item[(a.2)] Amongst all tree graphs, the chain and star graphs minimizes and maximizes, respectively, the information-theoretic lower bounds on the Bayes risk in  \eqref{eq:Bayes-Risk-LB-general} for a given fixed budget $n$, as defined in \eqref{eq:fixed-budget}.\footnote{Given that the chain graph topology is ``close'' to the ``optimal'' cycle graph topology, the optimality of chain graphs amongst trees is not surprising.}
\end{itemize}
Fig. \ref{MinimaxGraphs} illustrates the information-theoretic lower bounds as a function of the sample size in the discussed graph topologies.

\begin{figure}[t]
\begin{center}
\includegraphics[scale=0.45]{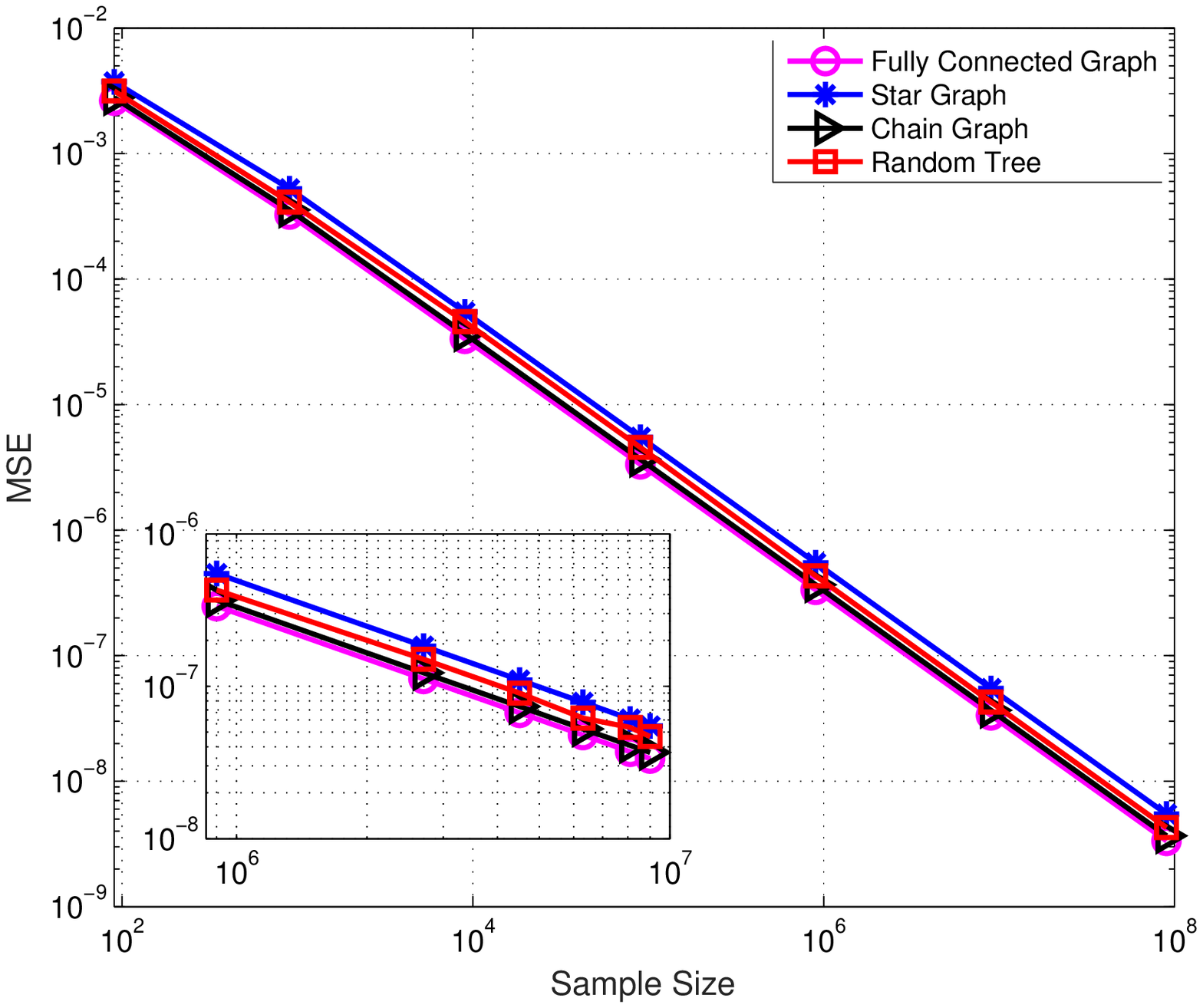}
\includegraphics[scale=0.45]{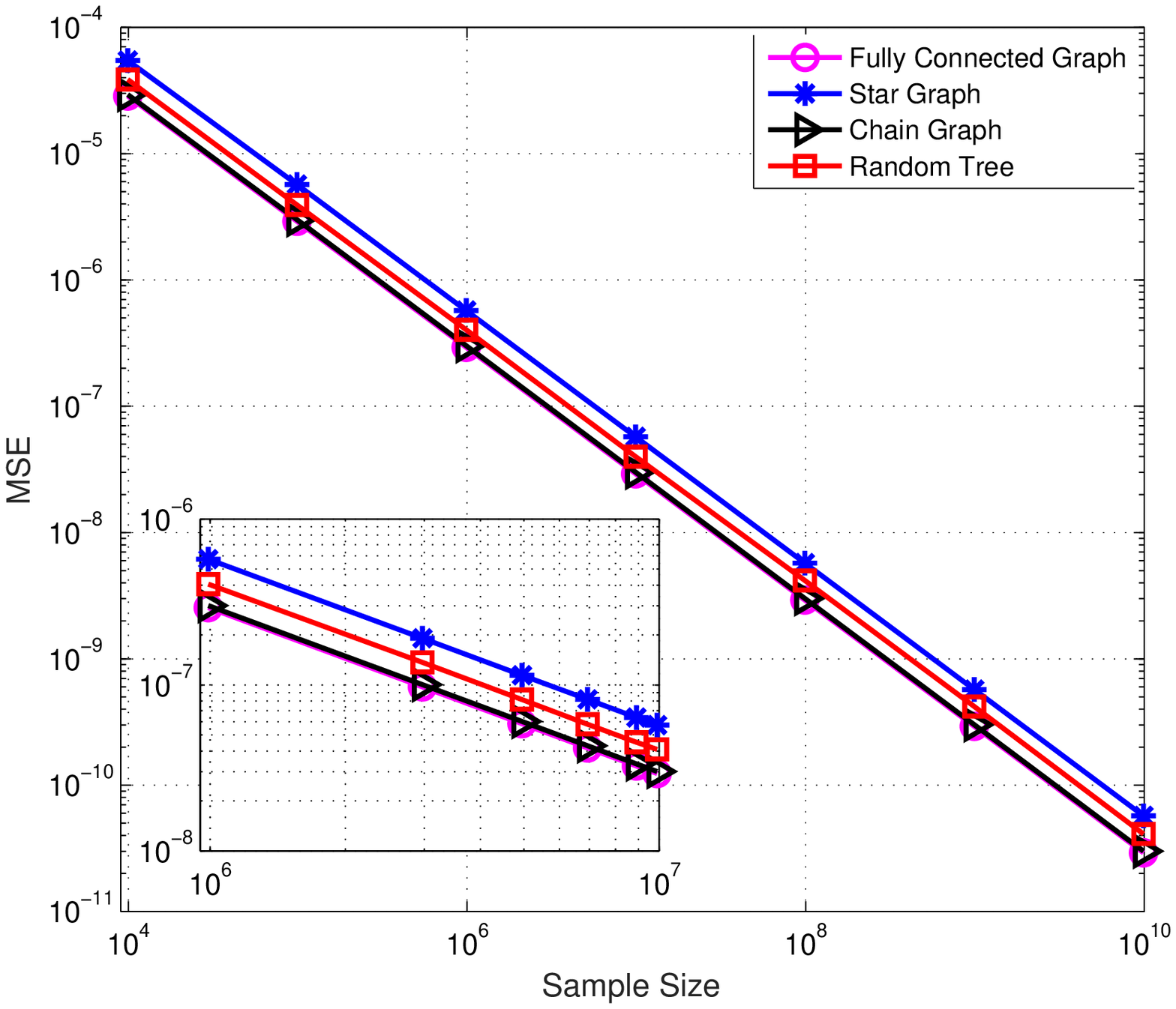}
\caption{BCRB as a function of number of samples for different graph topologies. Top figure is generated for $k=10$ items, and bottom figure for $k=100$ items. The parameters of the prior distribution in \eqref{eq:gamma-prior}  are chosen as $a=5$ and $b=ak-1$.} 
\label{BCRBGraphs}
\end{center}
\squeezeup
\end{figure}

Next, we analyze the dependence of the BCRB on graph topologies. Let $\matI_{\mathrm{st}}^{\bmLambda}$, $\matI_{\mathrm{ch}}^{\bmLambda}$,   $\matI_{\mathrm{ra}}^{\bmLambda}$, and  $\matI_{\mathrm{fc}}^{\bmLambda}$ denote the FIMs for a star graph, a chain graph, a random tree graph, and a fully connected graph, respectively. In Fig. \ref{BCRBGraphs}, we provide numerical evidence that for a given large budget $n$, as defined in \eqref{eq:fixed-budget}, the FIM of various graph topologies satisfy the inequalities:
\begin{equation}\label{eq:tr}
\textnormal{Tr}(({\matI_{\mathrm{fc}}^{\bmLambda}}))^{-1} \leq \textnormal{Tr}(({\matI_{\mathrm{ch}}^{\bmLambda}}))^{-1} 
\leq  \textnormal{Tr}(({\matI_{\mathrm{ra}}^{\bmLambda}}))^{-1} \leq  \textnormal{Tr}(({\matI_{\mathrm{st}}^{\bmLambda}}))^{-1}.
\end{equation}
Thus, we conjecture that the above answers (a.1) and (a.2) are also valid for the BCRB  when $n$ is large. A proof of this is left to future work.


\subsection{Phase Transitions}\label{subsec:phase-transitions}
\begin{figure}[t]
\squeezeup
\begin{center}
\includegraphics[scale=0.447]{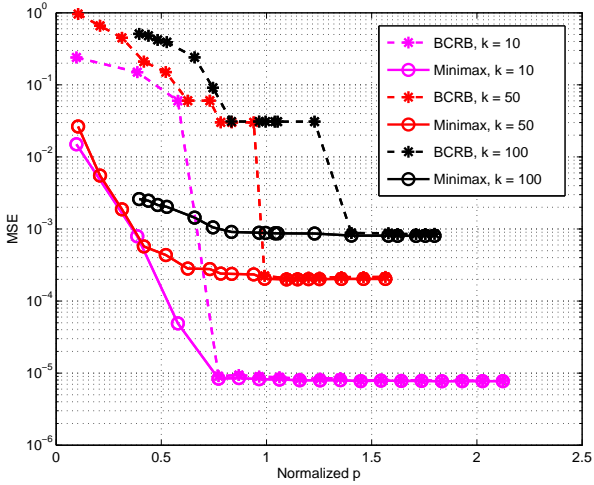}
\caption{Phase transition of the information-theoretic lower bound and the BCRB derived in Theorems \ref{thm:Bayes-Risk-LB} and \ref{thm:BCRB}, respectively, as a function of the normalized edge probability $p$ of the random ER graph different values of $k$ when $n$ is fixed.  The parameters of the prior in \eqref{eq:gamma-prior}  are chosen as $a=5$ and $b=25$.}
\label{BCRBMinimaxPhaseTrans}
\end{center}
\squeezeup
\end{figure}

To analyze the effect of graph connectedness on the derived lower bounds, we investigate whether our lower bounds demonstrate phase transitions as the number of edges increases. Let us assume that the edge set $E$ is drawn in accordance to the ER graph model where a node pair $(i,j)$ appears independently of any other node pair with probability $p\in(0,1)$.  We plot in  Fig.~\ref{BCRBMinimaxPhaseTrans} the information-theoretic lower bounds and the BCRBs as functions of the normalized edge probability of the random ER graph for various values of $k$ when $n$ is fixed. The edge probability $p$, which is given by the ratio of the non-zero edge weights over the total number of comparisons, is normalized by the factor $k^{-1}\log{k}$; this is because the phase transition for connectedness of an ER graph is given by the probability of edge appearance being  $k^{-1}\log{k}$. From the figure, we observe that the information-theoretic lower bounds derived in Theorem \ref{thm:Bayes-Risk-LB}  do not demonstrate sharp phase transitions, albeit a decrease is observed with increasing normalized edge probability. Thus, the bounds do not provide much information in terms of graph connectedness. On the other hand,  we notice that the BCRB derived in Theorem \ref{thm:BCRB} demonstrates a phase transition when the graph is almost connected corresponding to normalized probability $1$. This result might seem negative as phase transitions are useful to corroborate the validity of bounds in the sense that effective inference is not possible if ``the edge probability $<$ the critical threshold for connectedness''. However, the phase transition occurs in our model even when the graph may not be connected due to the inherent regularization present in the Bayesian nature of the problem. In particular, the priors allow for pairs of vertices $(i, j)$ to have $n_{ij}=0$ counts. 

\section{Extensions to the BTL Model with Home-Field Advantage}\label{sec:Home-Advantage}
It is reasonable to expect that in some applications, such as sport competitions, teams will have a better chance of winning when they play at home (compared to when they play in their opponent's home-field). The BTL model with home-field advantage \cite{CaronDoucet}  takes into account this asymmetry by associating to each item $i \in [k]$, a skill parameter $\lambda_i\in\mathbb{R}_+$ as before, but such that
\begin{equation}\label{eq:BTL-HA}
P_{ij}= 
\begin{cases}
Q_{ij} := \frac{\theta\lambda_i}{\theta\lambda_i + \lambda_j},& \text{if $i$ is home}, \\
\overline{Q}_{ij}  :=\frac{\lambda_i}{\lambda_i + \theta\lambda_j},  & \text{if $j$ is home},
\end{cases}
\end{equation}
where a new variable $\theta\in\mathbb{R}_{++}$ is introduced to model the strength of the home-field advantage ($\theta >1$) or disadvantage ($\theta < 1$).
Let $w_{ij}^{\mathrm{h}}$ denote the number of comparisons in which $i$ is at home and beats $j$. Let $n_{ij}^{\mathrm{h}}$ denote the total number of times $i$ and $j$ plays when $i$ is at home, so that $n_{ij} = n_{ij}^{\mathrm{h}} + n_{ji}^{\mathrm{h}}$. Note that the matrix  $\mathbf{N}^{\mathrm{h}}:=  (n_{ij}^{\mathrm{h}})\in\mathbb{N}^{k\times k}$  is not necessarily symmetric. As before, we assume that the total 
 budget matrix $\mathbf{N}:=  (n_{ij})\in\mathbb{N}^{k\times k}$ is fixed {\em a priori}. In this model, the data can be described by  $\mathbf{W}^{\mathrm{h}}:=  (w_{ij}^{\mathrm{h}})\in\mathbb{N}^{k\times k}$, and one can write 
\begin{equation}
\label{binomlikelihood}
p(\mathbf{W}^{\mathrm{h}}|\boldsymbol{\lambda}, \theta)\!=\!  \prod_{(i,j)\in\mathcal{I}_o[k]}\!\!\mathcal{B}(w_{ij}^{\mathrm{h}}; n_{ij}^{\mathrm{h}}, Q_{ij})\mathcal{B}(n_{ji}^{\mathrm{h}}-w_{ji}^{\mathrm{h}}; n_{ji}^{\mathrm{h}}, \overline{Q}_{ij}),
\end{equation}
by observing that $\Omega_{ij}^{\mathrm{h}}\sim \mathcal{B}(w_{ij}^{\mathrm{h}}; n_{ij}^{\mathrm{h}}, Q_{ij})$ holds for home-field wins, and 
$n_{ji}^{\mathrm{h}}-\Omega_{ji}^{\mathrm{h}}\sim\mathcal{B}(n_{ji}^{\mathrm{h}}-w_{ji}^{\mathrm{h}}; n_{ji}^{\mathrm{h}}, \overline{Q}_{ij})$  for foreign-field or away-field wins. As in the basic model, we assume that the skill parameter vector $\bmlambda$ follows the prior distribution given in \eqref{eq:gamma-prior}.   For this model, Caron and Doucet introduced the following latent variables~\cite[Eq.~(11)]{CaronDoucet}:
\begin{equation}\label{eq:Z-given-D-lambda-theta}
Z_{ij}^{\mathrm{h}} |\lambda_i, \lambda_j, \theta \sim p(\zeta_{ij}^{\mathrm{h}} |\lambda_i, \lambda_j, \theta) = \mathcal{G}(\zeta_{ij}^{\mathrm{h}}; n_{ij}^{\mathrm{h}}, \theta \lambda_i+\lambda_j),
\end{equation}
for all $(i, j)\in\mathcal{I}[k]$, and they showed that~\cite[eq.~(17)]{CaronDoucet} 
\begin{multline}\label{eq:lambda-given-D-Z-theta}
\Lambda_i| \mathbf{W}^{\mathrm{h}} , \boldsymbol{\zeta}^{\mathrm{h}} , \theta \sim p(\lambda_i| \mathbf{W}^{\mathrm{h}}, \boldsymbol{\zeta}^{\mathrm{h}}, \theta) \\*
= \mathcal{G}\Bigg(\lambda_i; a_i+\displaystyle\sum_{j\in[k]\setminus\{i\}}w_{ij}^{\mathrm{h}}+\displaystyle\sum_{j\in[k]\setminus\{i\}}\left(n_{ji}^{\mathrm{h}}-w_{ji}^{\mathrm{h}}\right), \\
  b_i+ \theta \displaystyle\sum_{j\in[k]\setminus\{i\}}\zeta_{ij}^{\mathrm{h}}  +  \displaystyle\sum_{j\in[k]\setminus\{i\}}\zeta_{ji}^{\mathrm{h}} \Bigg),
\end{multline}
for $i\in[k]$, where $\boldsymbol{\zeta}^{\mathrm{h}} =(\zeta_{ij}^{\mathrm{h}} )\in\mathbb{R}^{k\times k}$, As before, we use the symbols $\boldsymbol{\Omega}^{\mathrm{h}}:=(\Omega_{ij}^{\mathrm{h}})\in\mathbb{N}^{k\times k}$,  $\mathbf{Z}^{\mathrm{h}} :=(Z_{ij}^{\mathrm{h}} )\in\mathbb{R}^{k\times k}$, and $\Lambda:= (\Lambda_i)$ to denote the random matrices in the home-field advantage model, e.g., $\mathbf{\Omega}^{\mathrm{h}} $ refers to the data random variable with realizations given by $\mathbf{W}^{\mathrm{h}}$. Without loss of generality, we allow a prior distribution on the home-field advantage parameter such that $\Theta\sim p_\Theta(\theta)$, where $p_\Theta$ is a distribution with support  $(1,\infty)$. 

\subsection{Information-Theoretic Lower Bounds with Home-Field Advantage}\label{subsec:Home-Advantage-Minimax}
The next theorem provides a family of lower bounds obtained for the new model via Theorem \ref{thm:Raginsky}.  
\begin{thm}\label{thm:Bayes-Risk-LB-HA}
Consider the Bayesian BTL model with home-field advantage introduced in Section \ref{sec:Home-Advantage}. Let $\lVert \cdot \rVert$ denote an arbitrary norm in $\mathbb{R}^k$. For any  $r\geq 1$, let  $d(\boldsymbol{\lambda}, \boldsymbol{\widehat{\lambda}}) = \lVert \boldsymbol{\lambda} - \boldsymbol{\widehat{\lambda}}\rVert^r$ be the distortion function, where $\boldsymbol{\widehat{\lambda}}:= \boldsymbol{\varphi}(\mathbf{W}^{\mathrm{h}})$ is an estimator of $\boldsymbol{\lambda}$ based on data sample $\mathbf{W}^{\mathrm{h}}$ for a fixed $\mathbf{N}$. The Bayes risk $R_{\mathrm{B}}$ for estimating the parameter $\boldsymbol{\lambda}\in\mathbb{R}_{++}^k$ based on a sample $\mathbf{W}^{\mathrm{h}}$  in the Bayesian BTL model with home-field advantage is asymptotically lower bounded by the following expression:
\begin{align}
\label{eq:Bayes-Risk-LB-HA}R_{\mathrm{B}}& = \displaystyle\inf_{\boldsymbol{\varphi}} \Expt[d(\left(\boldsymbol{\Lambda},\Theta\right), \boldsymbol{\varphi}(\boldsymbol{\Omega}))]\nonumber\\
& \gtrsim_{n_i} \frac{k}{re}\left(V_k\Gamma\left(1+\frac{k}{r}\right)\right)^{-r/k} e^{-rE_{\mathrm{HA}}(\mathbf{N}^{\mathrm{h}}, \mathbf{a}, \mathbf{b},  p_\Theta ) } 
\end{align}
where $V_k$ denotes the volume of the unit ball in $(\mathbb{R}^k, \lVert\cdot\rVert)$, $n_i$ is defined in \eqref{eq:ni}, and 
\begin{align}\label{eq:E-HA}
&E_{\mathrm{HA}}(\mathbf{N}^{\mathrm{h}}, \mathbf{a}, \mathbf{b},  p_\Theta ) = \frac{1}{k}\sum_{i\in [k]} \Bigg( - \frac{1}{2}\log{(2\pi)} + \log{b_i} \nonumber \\
&-\!\psi(a_i)\!+\!\frac{1}{2}\log{\bigg(\!a_i \!+\!  \!\!\!\sum_{j\in[k]\setminus\{i\}}\! \!\! F_{ij}(n_{ij}^{\mathrm{h}}, n_{ji}^{\mathrm{h}}, a_i,b_i, p_\Theta ) \bigg)}\!\Bigg),
\end{align}
with 
\begin{align}
&F_{ij}(n_{ij}^{\mathrm{h}}, n_{ji}^{\mathrm{h}}, a_i,b_i, p_\Theta ) \nonumber\\
&\quad=  \Expt\left[\displaystyle\frac{\Theta\Lambda_i}{\Theta\Lambda_i+\Lambda_j}\right]n_{ij}^{\mathrm{h}} + \Expt\left[\displaystyle\frac{\Lambda_i}{\Lambda_i+\Theta\Lambda_j}\right]n_{ji}^{\mathrm{h}}  ,\label{eq:HA-theta-function}
\end{align}
for any $(i, j)\in\mathcal{I}[k]$.
\end{thm}

\begin{cor}
The lower bound in \eqref{eq:E-HA} justifies our basic intuition that one must choose $n_{ij}^{\mathrm{h}} = n_{ji}^{\mathrm{h}}$ to cancel the effect of any home-field advantage or disadvantage, since
\begin{equation}
 \Expt\left[\displaystyle\frac{\Lambda_i}{\Lambda_i+\Theta\Lambda_j}\right] =  \Expt\left[\displaystyle\frac{\Lambda_j}{\Lambda_j+\Theta\Lambda_i}\right] = 1 - 
\Expt\left[\displaystyle\frac{\Theta\Lambda_i}{\Theta\Lambda_i+\Lambda_j}\right].
\end{equation}
  Thus, symmetric matrices $\mathbf{N}^{\mathrm{h}}$ lead to $E_{\mathrm{HA}}(\mathbf{N}^{\mathrm{h}}, \mathbf{a}, \mathbf{b},  p_\Theta ) = E_{\mathrm{BTL}}(\mathbf{N}, \mathbf{a}, \mathbf{b})$, which is given by \eqref{eq:E1}.
\end{cor}

\begin{figure}[t]
\squeezeup
\begin{center}
        \includegraphics[scale=0.3]{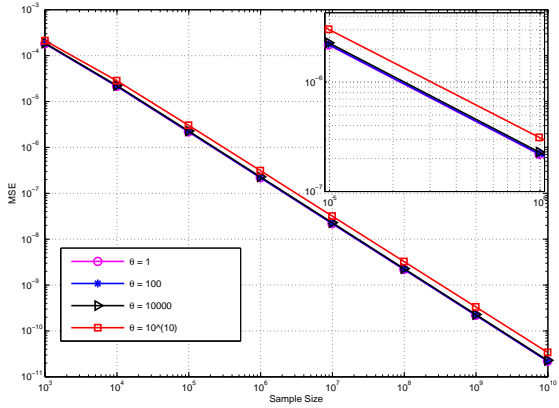}
                  \caption{Impact of the home-field advantage parameter $\theta>1$ on the information-theoretic lower bounds of Theorem \ref{thm:Bayes-Risk-LB-HA} for the squared $L^2$  norm. The figure is generated based on $k=10$ items and for the case $n_{ij}^{\mathrm{h}} = n_{ij}$, for $(i, j)\in\mathcal{I}_o[k]$. The parameters of the prior distribution in \eqref{eq:gamma-prior}  are chosen as $a=2$ and $b=ak-1$.} 
                   \label{MinimaxHA}
\end{center}
\squeezeup
\end{figure}

Suppose that the symmetry condition is not satisfied, i.e., $n_{ij}^{\mathrm{h}} \neq n_{ji}^{\mathrm{h}}$ holds for some pairs of items $(i, j)\in\mathcal{I}_o[k]$. 
In this case, we want to analyze how the home-field advantage parameter affects the family of information-theoretic  lower bounds in \eqref{eq:Bayes-Risk-LB-HA}. For this purpose, we now discuss a special case of Theorem \ref{thm:Bayes-Risk-LB-HA}, where we evaluate  \eqref{eq:HA-theta-function} by symbolic computing software for deterministic $\Theta=\theta > 1$ and constant $a_i = a$ and $b_i=b$, for all $i\in[k]$. In this case, we get
\begin{align}
&\Expt\left[\displaystyle\frac{\theta\Lambda_i}{\theta\Lambda_i+\Lambda_j}\right]  = f(a, \theta) \nonumber\\
&\qquad:= a \left(-1+\frac{1}{\theta}\right)^{-2 a} \theta^{-a} B[1-\theta,2 a,1-a],\label{eq:HA-theta-function-deterministic}
\end{align}
where $B[z, x, y]$ is the incomplete beta function \cite{IBF}. Therefore, we see that  \eqref{eq:HA-theta-function} does not actually depend on the scale parameter $b$ of the Gamma prior in \eqref{eq:gamma-prior} (and this is true for both random and deterministic $\Theta$).  Moreover, it can be verified that $\lim_{\theta\to1} f(a, \theta) = 1/2$ holds as expected, and  $\lim_{\theta\to\infty} f(a, \theta) = 1$ . 
In particular, for $a=2$, \eqref{eq:HA-theta-function-deterministic} reduces to the following simpler form
\begin{equation}
f(2, \theta)= \displaystyle\frac{\theta (2+3 \theta-6 \theta^2+\theta^3+6 \theta \log{\theta})}{(-1+\theta)^4}.
\end{equation}
It can be verified that function  $f(2, \theta)$ is increasing and concave if $\theta > 1$, for any $a\in\mathbb{R}_{++}$. Moreover, $f(10) \approx 0.87$ and $f(100) \approx 0.98$. Fig. \ref{MinimaxHA} illustrates the impact of the parameter $\theta> 1$ on the lower bounds in \eqref{eq:Bayes-Risk-LB-HA} for a particular choice of the matrix $\mathbf{N}^{\mathrm{h}}$ for $k=10$ items. In fact, letting $n_{ij}^{\mathrm{h}} = \alpha n_{ij}$, for $(i, j)\in\mathcal{I}_o[k]$ and $\alpha\in(0.5, 1)$, \eqref{eq:HA-theta-function} equals $\left((2\alpha-1)f(2,\theta) + (1-\alpha)\right)n_{ij}$, for $(i, j)\in\mathcal{I}_o[k]$, and $\alpha n_{ij}-(2\alpha-1)f(2, \theta)n_{ij}$, for $(i,j)\in\mathcal{I}[k]\setminus\mathcal{I}_o[k]$. The observed behavior in Fig. \ref{MinimaxHA}  can be better understood by inspecting the latter relations.

The proof  of Theorem \ref{thm:Bayes-Risk-LB-HA} relies on Theorem~\ref{thm:Raginsky} and the following proposition proved in the supplementary material~\cite{supp}.
\begin{prop}\label{prop:MI-HA}
We have
\squeezeup
\begin{equation}\label{eq:MI-HA}
\frac{1}{k}\left(I(\boldsymbol{\Lambda}; \boldsymbol{\Omega}\mathbf{Z}) - h(\boldsymbol{\Lambda})\right)   \lesssim_{n_i}  E_{\mathrm{HA}}(\mathbf{N}^{\mathrm{h}}, \mathbf{a}, \mathbf{b},  p_\Theta ),
\squeezeup
\end{equation}
where $n_i$ is defined in \eqref{eq:ni} and $E_{\mathrm{HA}}(\mathbf{N}^{\mathrm{h}}, \mathbf{a}, \mathbf{b},  p_\Theta )$ in \eqref{eq:E-HA}.
\end{prop}
We omit the proof of Theorem \ref{thm:Bayes-Risk-LB-HA} since it is proved using similar steps to the proof of Theorem  \ref{thm:Bayes-Risk-LB}.

\subsection{Hybrid Cram\'{e}r-Rao Lower Bounds with Home-Field Advantage}\label{subsec:Home-Advantage-BCRB}
We derive the HCRB for the BTL model with home-field advantage described in~\eqref{eq:BTL-HA}. The Cram\'{e}r-Rao bound derived here is \emph{hybrid} as it is obtained using the HIM computed over the random  vector $\bmLambda$ and the deterministic parameter $\theta > 1$.  The likelihood is given by \cite{CaronDoucet}
\begin{align}\label{eq:LikelihoodHCRBHA}
p(\matD^{ \mathrm{h}}, \bmlambda| \theta) &= \prod_{i \in[k]} \frac{b^{a_i}}{\Gamma(a_i)} \lambda_i^{a_i} e^{-b\lambda_i} \nonumber\\
&\!\!\!\!\!\!\!\!\!\!\!\!\!\!\!\!\!\!\times\prod_{(i, j)\in\mathcal{I}[k]} {n_{ij}^\mathrm{h} \choose w_{ij}^\mathrm{h}} \left(\frac{\theta\lambda_i}{\theta\lambda_i + \lambda_j}\right)^{w_{ij}^\mathrm{h}} \left(\frac{\lambda_j}{\theta\lambda_i + \lambda_j}\right)^{n_{ij}^\mathrm{h}-w_{ij}^\mathrm{h}},
\end{align}
where we recall that $w_{ij}^{\mathrm{h}}$ denote the number of comparisons in which $i$ is at home and beats $j$, and $n_{ij}^{\mathrm{h}}$ denote the total number of times $i$ and $j$ plays when $i$ is at home, for all $(i, j)\in\mathcal{I}[k]$.  In the following, we state the HCRB.
\begin{thm}\label{thm:Bayes-Risk-BCRB-HA}
Consider the Bayesian BTL model with home-field advantage introduced in Section \ref{sec:Home-Advantage}. Define the expectations of $\frac{\Lambda^{t_i}_i\Lambda_j^{t_j}}{\theta\Lambda_i+\Lambda_j}$ and  $ \frac{1}{(\theta\Lambda_i+\Lambda_j)^2}$ for $t_i,t_j \in (-\infty,\infty)$ respectively as 
\begin{align}
\label{eq:expt-1-nsf}\mu_{\Lambda_i\Lambda_j} (t_i,t_j,\theta ) &:= \Expt\left[ \frac{\Lambda^{t_i}_i\Lambda_j^{t_j}}{\theta\Lambda_i+\Lambda_j}\right]  \\
\label{eq:expt-2-nsf}\nu_{\Lambda_i,\Lambda_j}(t_i,t_j,\theta) & := \Expt\left[ \frac{1}{(\theta\Lambda_i+\Lambda_j)^2}\right].
\end{align}
Given the joint probability distribution   in \eqref{eq:LikelihoodHCRBHA}, the HCRB  on the MSE matrix $\mathbf{E}^{\bmlambda,\theta}$ of the unknown hybrid vector $[\bmlambda, \theta]$, where the home-field advantage parameter $\theta$ is deterministic, is given by $\mathbf{E}^{\bmlambda, \theta} \succeq ({\mathbf{I}_{\mathrm{HA}}^{\bmlambda, \theta}})^{-1}$, where
\begin{align}
&\mathbf{I}_{\mathrm{HA}}^{\bmlambda,\theta} :=
\begin{bmatrix}
\mathbf{H}^{\bmlambda} &   \mathbf{H}^{\bmlambda,\theta}\\
(\mathbf{H}^{\bmlambda,\theta})^T & \mathbf{H}^{\theta}
\end{bmatrix}
\end{align}
such that 
\begin{align}
\label{eq:eq-bcrb-ha-1}&[\matH^{\bmLambda}]_{i,i} := \frac{(a_i-1)b^2\Gamma(a_i-2)}{\Gamma(a_i)}\nonumber\\
& \quad+  \sum_{j\in[k]\setminus\{i\}} n_{ij}^{\mathrm{h}} \theta \nu_{\Lambda_i,\Lambda_j}(- 1,1,\theta) \nonumber\\
& \quad+   \sum_{j\in[k]\setminus\{i\}} n_{ji}^{\mathrm{h}} \theta \nu_{\Lambda_j,\Lambda_i}(1,- 1,\theta) , ~ \forall i\in[k]\\
\label{eq:eq-bcrb-ha-2}&[\matH^{\bmLambda}]_{i,j}:=  -\left[ n_{ij}^{\mathrm{h}} \theta \nu_{\Lambda_i,\Lambda_j}(- 1,1,\theta) \right.\nonumber\\
&\left.+  n_{ji}^{\mathrm{h}} \theta \nu_{\Lambda_j,\Lambda_i}( 1,- 1,\theta) \right],~ ~ \forall (i, j)\in\mathcal{I}[k], \\
\label{eq:eq-bcrb-ha-3}&[\matH^{\theta}]_{1,1} :=   \sum_{(i,j)\in\mathcal[k]}   \frac{n_{ij}^{\mathrm{h}}}{\theta} \mu_{\Lambda_i\Lambda_j}(0,0,\theta)\nonumber\\
&\qquad\qquad-  \sum_{(i,j)\in\mathcal[k]}  n^{\mathrm{h}}_{ij}  \nu_{\Lambda_i,\Lambda_j}(2,0,\theta),\\
\label{eq:eq-bcrb-ha-4}&[\matH^{\bmLambda,\theta}]_{i,1} :=  \sum_{j\in[k]\setminus\{i\}}\left[n_{ij}^{\mathrm{h}} \mu_{\Lambda_i\Lambda_j}(-1,0,\theta) \right.\nonumber\\
&\left.\quad- n_{ij}^{\mathrm{h}}\theta\nu_{\Lambda_i,\Lambda_j}(1,0,\theta)\! -\!  n_{ji}^{\mathrm{h}}\theta\nu_{\Lambda_j,\Lambda_i}(1,0,\theta)\right] \nonumber\\
&\qquad \qquad \qquad~~ \forall i\in[k],
\end{align}
where the expressions for the quantities  $\mu_{\Lambda_i,\Lambda_j}(t_i,t_j,\theta)$ and $\nu_{\Lambda_i,\Lambda_j}(t_i,t_j,\theta) $ are provided in Lemmas 5 and 6 in  the supplementary material~\cite{supp}.
\end{thm}

In Section \ref{subsec:Main-Results-Bayesian-CRB}, we saw that the BCRB computation involves obtaining the mean of  $\frac{\Lambda_i}{\Lambda_i + \Lambda_j}$ w.r.t.\ $\Lambda_i$ and $\Lambda_j$, which  is  straightforward. However, due to the presence of the parameter $\theta>1$, deriving the expressions of the mean of $\frac{\theta\Lambda_i}{\theta\Lambda_i + \Lambda_j}$ is not  straightforward. We derive this mean and generalize it to obtain the expressions for $\mu_{\Lambda_i,\Lambda_j}(t_i,t_j,\theta)$ and $\nu_{\Lambda_i,\Lambda_j}(t_i,t_j,\theta) $ in the supplementary material.

\section{Conclusions}\label{sec:Conclusions}
We presented two families of lower bounds on the Bayes risk  for learning the skill parameters of the Bayesian BTL model $\boldsymbol{\lambda}$. From these bounds, we made progress in understanding the effect of the various graph structures (indicating the pairs of items who are compared against one another) on the Bayes risk of the Bayesian BTL model.

There are multiple directions for future research. First, we would like to assess the tightness of the derived lower bounds by deriving matching upper bounds. From  Fig.~\ref{fig:MSE_EM_BCRB_Raginsky}, it appears that the bounds are increasingly tight as the sample size $n\to\infty$. Showing that this is true analytically would be of tremendous theoretical interest and would confirm that the answers to the questions we posed in the Introduction are based not only on lower but also on upper bounds. Second, we would like to show that \eqref{eq:tr} is true, which would  imply  that the BCRB allows us to make the same conclusions on graph structures as the family of information-theoretic lower bounds. Finally, we would like to use the bounds to gain further intuition on how  the structure  of the comparison graph affects the minimax risk. Some questions of interest include: Does the fully-connected graph outperform a simple cycle (this was left unexplored in answer (a.1))? For a fixed number of edges, do planar graphs generally outperform non-planar ones?

\appendices
\section{Proof of Proposition \ref{prop:MI}} \label{app:proofProp}
\begin{IEEEproof}
We first note that 
\begin{equation}\label{eq:MI}
I(\boldsymbol{\Lambda};  \boldsymbol{\Omega},\mathbf{Z})  =  \Expt\left[\log{\displaystyle\frac{p(\boldsymbol{\Lambda},  \boldsymbol{\Omega},\mathbf{Z})}{p(\boldsymbol{\Lambda})p( \boldsymbol{\Omega},\mathbf{Z})}}\right] 
= \Expt\left[\log{\displaystyle\frac{p(\boldsymbol{\Lambda}| \boldsymbol{\Omega},\mathbf{Z})}{p(\boldsymbol{\Lambda})}}\right] .
  \end{equation}
 Using the last expression, it is easy to see that we have
  \begin{equation}\label{eq:exponent}
I(\boldsymbol{\Lambda};  \boldsymbol{\Omega},\mathbf{Z}) - h(\boldsymbol{\Lambda}) = \Expt\left[\log{p(\boldsymbol{\Lambda}| \boldsymbol{\Omega},\mathbf{Z})}\right].
  \end{equation}
On the other hand, by Lemma 3 given in the supplementary material~\cite{supp},  we know that the skill parameters of the Bayesian BTL model follow the following conditional probability distribution:
  \begin{equation}
p(\boldsymbol{\lambda}|\mathbf{W},\boldsymbol{\zeta}) = \displaystyle\prod_{i\in[k]}\mathcal{G}(\lambda_i; a_i+w_i, b_i+\zeta_i),
 \end{equation}
 where $w_i$ is the total number of wins of an item $i$ given by
$w_i := \sum_{j\in[k]\setminus\{i\}} w_{ij}$ and
$\zeta_i := \sum_{j\in[k]\setminus\{i\}} \zeta_{ij}$, 
for all $i\in[k]$. The random variables corresponding to these realizations are denoted as $\Omega_i$ and $Z_i$, respectively.  Thus, to prove \eqref{eq:MI-Bound}. we need to compute an upper bound to
 \begin{equation}
I(\boldsymbol{\Lambda}; \boldsymbol{\Omega},\mathbf{Z}) - h(\boldsymbol{\Lambda})  =\sum_{i\in[k]}\Expt\left[\log\mathcal{G}(\Lambda_i; a_i+\Omega_i, b_i+Z_i)\right].
\end{equation}
For that purpose, we first claim that
\begin{equation}
\lim_{n_i\to\infty} \log{\left(1+O\left(\Expt\left[\frac{1}{a_i + \Omega_i }\right]\right)\right)} = 0,
\end{equation}
where $n_i$ is defined in \eqref{eq:ni}. For the proof, see Lemma 4 in the supplementary material~\cite{supp}. 
Now, using the identifications $M\leftarrow \Lambda_i$, $A \leftarrow a_i+\Omega_i$, and $B\leftarrow  b_i+Z_i$,  we get by Proposition \ref{prop:Expt-general} presented at the end of  this Appendix, the following asymptotic upper bound:
\begin{multline}\label{eq:terms}
I(\boldsymbol{\Lambda}; \boldsymbol{\Omega},\mathbf{Z}) - h(\boldsymbol{\Lambda}) 
\le\displaystyle\sum_{i\in[k]} \Bigg( - \frac{1}{2}\log{(2\pi)} - \Expt\left[\log  \Lambda_i \right] \\
+ \frac{1}{2}\log{\left(a_i+\Expt\left[\Omega_i\right]\right)}  + \Expt\left[(a_i+\Omega_i)-(b_i+Z_i)\Lambda_i\right]
\\  +  \log{\left(1+O\left(\Expt\left[\frac{1}{a_i + \Omega_i }\right]\right)\right)}\Bigg),
\end{multline}
as $n_i\to\infty$. 
 We are only left to compute the terms in~\eqref{eq:terms}. We start by computing
\begin{align}
\Expt\left[a_i+\Omega_i\right] &= \Expt\left[a_i + \displaystyle\sum_{j\in[k]\setminus\{i\}} \Omega_{ij} \right] \\
&= a_i + \displaystyle\sum_{j\in[k]\setminus\{i\}}  \Expt\left[\Expt\left[\Omega_{ij}|\Lambda_i, \Lambda_j\right] \right] \\
\label{eq:t1}&=a_i +  \displaystyle\sum_{j\in[k]\setminus\{i\}} \Expt\left[n_{ij}\displaystyle\frac{\Lambda_i}{\Lambda_i+\Lambda_j} \right],
\end{align}
 where \eqref{eq:t1} follows from $\Omega_{ij}|\lambda_i, \lambda_j \sim  \mathcal{B}(w_{ij}; n_{ij}, P_{ij})$.  Next, we compute 
  \begin{align}
\Expt\left[(b_i+Z_i)\Lambda_i\right]   &=   \Expt\left[ \Bigg(b_i + \displaystyle\sum_{j\in[k]\setminus\{i\}}  Z_{ji}\Bigg)\Lambda_i\right]  \\
  &= b_i \Expt\left[\Lambda_i\right]  +   \displaystyle\sum_{j\in[k]\setminus\{i\}} \Expt\left[Z_{ji}\Lambda_i\right]\\
  &= b_i \Expt\left[\Lambda_i\right]  +   \displaystyle\sum_{j\in[k]\setminus\{i\}} \Expt\left[\Lambda_i\Expt\left[Z_{ji}|\Lambda_i, \Lambda_j\right]\right]\\
  &=   b_i\frac{a_i}{b_i}+  \displaystyle\sum_{j\in[k]\setminus\{i\}} \Expt\left[ \Lambda_i\frac{n_{ij}}{\Lambda_i+\Lambda_j}\right]  \label{eq:t2-} \\
 \label{eq:t2} &= a_i +  \displaystyle\sum_{j\in[k]\setminus\{i\}}\Expt\left[\displaystyle\frac{n_{ij} \Lambda_i}{\Lambda_i+\Lambda_j}\right]  
 \end{align}
 where \eqref{eq:t2-} follows from the fact that $Z_{ji} |\lambda_i,\lambda_j \sim \mathcal{G}(\zeta_{ij}; n_{ij}, \lambda_i+\lambda_j)$ and $\Lambda_i\sim\mathcal{G}(\lambda_i, a_i, b_i)$. Thus, we conclude from \eqref{eq:t1} and \eqref{eq:t2} that the two terms cancel, i.e.,
$\Expt\left[\left(a_i+\Omega_i\right) - \left(b_i+Z_i\right)\Lambda_i\right]= 0$.  Finally, the proof is completed by noting that for $\Lambda_i\sim\mathcal{G}(\lambda_i, a_i, b_i)$, we have $ \Expt\left[\log{\Lambda_i}\right] = \psi(a_i)-\log{b_i}$ \cite{gammaDist}, and for $\Omega_{ij}|\lambda_i, \lambda_j \sim  \mathcal{B}(w_{ij}; n_{ij}, P_{ij})$, we have
\begin{equation}
\frac{1}{2}\log{\left(a_i+\Expt\left[\Omega_i\right]\right)} = \frac{1}{2}\log{\Bigg( a_i + \frac{1}{2}\displaystyle\sum_{j\in[k]\setminus\{i\}}n_{ij}\Bigg)}.
\end{equation} 
\end{IEEEproof}

\begin{prop}\label{prop:Expt-general}
Let $M$, $A$ and $B$ be three non-negative random variables for which we define the random variable 
$\mathcal{G}(M; A, B)$, where $A$ and $B$ determines respectively the shape and rate parameters of a random Gamma distribution of $M$. Then,  as $\Expt\left[1/A\right]\to0$,
\begin{multline}\label{eq:Expt-gamma-approx}
\Expt\left[\log\mathcal{G}(M; A, B)\right] \le  - \frac{1}{2}\log{(2\pi)} - \Expt\left[\log M\right] \\
+ \frac{1}{2}\log{ \Expt\left[A\right] }  \!+\!\Expt\left[A\!-\! BM\right] 
\!+\! \log{\left(1\! +\!O\left(\Expt\left[\frac{1}{ A}\right]\right)\right)}.
\end{multline}
\end{prop}
\begin{IEEEproof}
We start by writing
\begin{multline}\label{eq:log-gamma}
\log{\mathcal{G}(M; A, B)} = \log\left(\displaystyle\frac{B^{A}}{\Gamma(A)}M^{ A-1}e^{-BM}\right)\\
=A\log  B -\log \Gamma(A)+\left(A-1\right)\log   M -BM.
\end{multline}
As the Gamma function can be approximated using Stirling's formula \cite{Stirling}, i.e., 
 \begin{multline}
 \log{   \Gamma(x)}  = \frac{1}{2}\log{(2\pi)}+ x\log{x} -\frac{1}{2}\log{x} - x \\
 + \log{\left(1+O\left(\frac{1}{x}\right)\right)}
\end{multline}
holds for any $x\in\mathbb{R}$, we obtain the following asymptotic expression for \eqref{eq:log-gamma}:
\begin{multline}\label{eq:gamma-approx}
\log{\mathcal{G}(M; A, B)} = A\log  B  +\left(A-1\right)\log M -BM \\
- \bigg( \frac{1}{2}\log{(2\pi)}+  A\log{A} -\frac{1}{2}\log{A} -  A  \\
+ \log{\left(1+O\left(\frac{1}{ A}\right)\right)}\bigg).
\end{multline}
As a result, to prove the claim in \eqref{eq:Expt-gamma-approx}, we compute an upper bound on  $\Expt\left[\log\mathcal{G}(M; A, B)\right]$ using the approximation in \eqref{eq:gamma-approx}. 
First, we show that 
 \begin{equation}\label{eq:relation-2}
\Expt\left[ A \log\left( BM\right) - A \log A \right] \leq 0.
\end{equation}
To prove this claim, we write
\begin{align}
 &\Expt\left[ A \log\left( BM\right) - A \log A  \right] \\
 &= \Expt\left[ A \Expt\left[\log\left(BM\right)|B\right] -  A\log{ A} \right]\\
\label{eq:i1} &\leq \Expt\left[ A\log\left(\Expt\left[BM|B\right]\right) -  A\log{ A} \right]\\
&=\Expt\left[ A\log\left(B\Expt\left[M|AB\right]\right) -  A\log{ A} \right]\\
\label{eq:i2} &\leq  \Expt\left[ A\log\left(B\displaystyle\frac{A}{B}\right) -  A\log{ A} \right]= 0,
 \end{align}
 where \eqref{eq:i1} follows by Jensen's inequality for concave functions, and \eqref{eq:i2} follows by the fact that 
$\Expt\left[M|AB\right] = A/B$ holds for the Gamma distribution \cite{gammaDist}. 
Next, by Jensen's inequality, 
\begin{equation}\label{eq:relation-3}
 \Expt\left[\frac{1}{2}\log A\right] \leq \frac{1}{2}\log  \Expt\left[A\right],
 \end{equation} 
and
\begin{equation}\label{eq:relation-4}
 \Expt\left[ \log{\left(1+O\left(\frac{1}{ A}\right)\right)}\right] \leq\log{\left(1+O\left(\Expt\left[\frac{1}{ A}\right]\right)\right)}.
 \end{equation} 
 By upper bounding $\Expt\left[\log\mathcal{G}(M; A, B)\right]$ via \eqref{eq:relation-2}, \eqref{eq:relation-3}, and \eqref{eq:relation-4}, we obtain the claim of the lemma.
\end{IEEEproof}

\section{Proof of Theorem \ref{thm:BCRB}} \label{app:prf_bcrb}

\begin{IEEEproof}
Using the BTL model given in Section \ref{subsec:Pre-Bayesian-BTL}, the log-likelihood is given by
\begin{multline}\label{eq:loglikeli}
\log p(\matD,\bmlambda) =  \sum_{(i, j)\in\mathcal{I}_0[k]} \log {n_{ij} \choose w_{ij}} \\
+ \sum_{(i, j)\in\mathcal{I}[k]} \left[ w_{ij} \log(\lambda_i) - w_{ij} \log(\lambda_i + \lambda_j)\right] \\
+ \sum_{i\in[k]} \left[ a_i\log b - \log \Gamma(a_i) + (a_i-1) \log \lambda_i  - b \lambda_i\right].
\end{multline}
Differentiating \eqref{eq:loglikeli} w.r.t.\ $\lambda_i$, we obtain
\begin{align}
&\frac{\partial \log p(\matD,\bmlambda)}{\partial \lambda_i} \nonumber\\
&=  \frac{ a_i-1 + \sum_{j = 1}^{k} w_{ij} }{\lambda_i} 
- \sum_{j\in[k]\setminus\{i\}} \left(\frac{w_{ij}}{ \lambda_i + \lambda_j } +\frac{w_{ji}}{ \lambda_i + \lambda_j }\right) \\
\label{eq:diff1ii}&= \frac{ a_i-1 + \sum_{j = 1}^{k} w_{ij} }{\lambda_i} - \sum_{j\in[k]\setminus\{i\}}\frac{n_{ij}}{ \lambda_i + \lambda_j } ,
\end{align}
 for $i\in[k]$, where we used the fact that $n_{ij} = w_{ij}+w_{ji}$ holds for all $(i, j)\in\mathcal{I}[k]$. Differentiating \eqref{eq:diff1ii} w.r.t.\ $\lambda_i$, we obtain
\begin{multline}\label{eq:diff2ii}
\frac{\partial^2 \log p(\matD,\bmlambda)}{\partial \lambda_i^2} \\
= -\frac{(a_i-1)+\sum_{j = 1}^k w_{ij}}{\lambda_i^2} +   \sum_{j\in[k]\setminus\{i\}} \frac{n_{ij}} {(\lambda_i+\lambda_j)^2},
\end{multline}
for $i\in[k]$. Differentiating \eqref{eq:loglikeli} w.r.t.\ $\lambda_i$ and $\lambda_j$ we get
\begin{equation}
\frac{\partial^2 \log p(\matD,\bmlambda)}{\partial \lambda_i \partial \lambda_j} =   \frac{n_{ij}}{(\lambda_i + \lambda_j)^2},
\label{eq:diff2ij}
\end{equation}
 for $(i, j)\in\mathcal{I}[k]$. In order to obtain the BCRB, we take the expectations of \eqref{eq:diff2ii} and \eqref{eq:diff2ij} w.r.t.\ the joint density function. Since $\Expt\left[\Omega_{ij}|\Lambda_i, \Lambda_j\right]= \frac{n_{ij}\Lambda_i}{\Lambda_i + \Lambda_j}$,  we have
\begin{equation}
 [ \matI^{\bmLambda}]_{i,i}  =\Ex\Bigg[  \frac{1-  a_i  }{\Lambda_i^2} 
+  \sum_{j\in[k]\setminus\{i\}} \!\frac{{n_{ij}}\Lambda_j}{\Lambda_i(\Lambda_i+\Lambda_j)^2}\Bigg].
\end{equation}
Evaluating the above expression  we get \eqref{eq:BIM-ii}. Furthermore, we compute the off-diagonal terms as
\begin{equation}
 [\matI^{\bmLambda}]_{i,j}= - n_{ij} T_3(a_i,a_j,b),
\end{equation}
for $(i, j)\in\mathcal{I}_[k]$. To obtain  an expression for the BCRB, we are only left to compute the expressions for $T_1$, $T_2$, and $T_3$ given by \eqref{eq:T1}, \eqref{eq:T2}, and \eqref{eq:T3}, respectively. It is easy to see that $T_1(a_i,b)$ is given by \eqref{eq:T1}. We compute $T_3(a_i,a_j,b)$ as
\begin{multline}
 T_3(a_i,a_j,b) = \Ex\left[\frac{1}{(\Lambda_i + \Lambda_j)^2} \right]\label{eq:T3a}\\
= c_{\bmlambda}\int_{\lambda_i} \left\{\int_{\lambda_j} \frac{1}{(\lambda_i + \lambda_j)^2} \lambda_j^{a_j-1} e^{-b\lambda_j} \mathrm{d}\lambda_j\right\}\lambda_i^{a_i-1} e^{-b\lambda_i} \mathrm{d}\lambda_i,\nonumber
\end{multline}
where $c_{\bmlambda} = \frac{b^{(a_i+a_j)}}{\Gamma(a_i)\Gamma(a_j)}$. We first compute the integral given by
\begin{equation}
I_3(\lambda_i,a_j,b) = \int_{\lambda_j = 0}^{\infty} \frac{\lambda_j^{a_j-1} e^{-b\lambda_j}}{(\lambda_i + \lambda_j)^2}  \mathrm{d}\lambda_j.
\end{equation}
Using integration by parts, we obtain  
\begin{multline}
 T_3(a_i,a_j,b)  = \\
c_{\bmlambda}(a_j \!-\! 1) \int_{\lambda_i} \int_{\lambda_j}\frac{\lambda_i}{ \lambda_i+\lambda_j }\lambda_i^{ a_i-2 }e^{-b\lambda_i} \lambda_j^{ a_j-2 }e^{-b\lambda_j}\, \mathrm{d}\lambda_i\, \mathrm{d}\lambda_j \\
- c_{\bmlambda}b \int_{\lambda_i} \int_{\lambda_j}\frac{\lambda_i}{ \lambda_i+\lambda_j }\lambda_i^{ a_i-2 }e^{-b\lambda_i} \lambda_j^{ a_j-1 }e^{-b\lambda_j} \, \mathrm{d}\lambda_i \, \mathrm{d}\lambda_j,\nonumber
\end{multline}
where we apply the limits  $\left.\frac{-\lambda_j^{(a_j-1)}e^{-b\lambda_j}}{ \lambda_i+\lambda_j }\right]_{\lambda_j = 0}^{\infty} = 0$. It is well-known that if $X \sim \mathcal{G}(x;\alpha_x,\beta)$ and $Y \sim \mathcal{G}(y;\alpha_y,\beta)$, then $\frac{X}{X+Y} \sim \mathrm{Beta}(\alpha_x,\alpha_y)$  and hence, $\Ex\left[\frac{X}{X+Y}\right] = \frac{\alpha_x}{\alpha_x+\alpha_y}$. Using this result for each term in $T_3(a_i,a_j,b)$, we obtain the expression in \eqref{eq:T3}. For integer values of $a_i$ and $a_j$, we get $T_3(a_i,a_j,b) = b^2/\left((a_i+a_j-1)(a_i+a_j-2)\right)$.
Further, $T_2(a_i,a_j,b)$ is given by
\begin{equation}
T_2(a_i,a_j,b) = \Ex\left[\frac{\Lambda_j}{\Lambda_i(\Lambda_i+\Lambda_j)^2}\right].
\end{equation}
Using the techniques to simplify $T_3(a_i,b)$, we obtain $T_2(a_i,b)$ as in \eqref{eq:T2}. For integer values of $a_i$ and $a_j$, we obtain 
$T_2(a_i,a_j,b) =  b^2 a_j/\left((a_i+a_j - 1)(a_i+a_j - 2)\right)$.
\end{IEEEproof}

\section{Proof of Corollary \ref{cor:Minimax-optimal-tree}} \label{app:prf_tree}
\begin{IEEEproof} 
 Let us first prove the claim concerning the star graph. We first note that, for a fixed $n$ as in \eqref{eq:fixed-budget},  maximizing the lower bounds on the Bayes risk in  \eqref{eq:Bayes-Risk-LB-general} is equivalent to minimizing the following sum
\begin{align}\label{eq:star-sum-rate}
S:=\frac{1}{2}\log{\Bigg(a + 2n -  \sum_{i'\in[k]\setminus\{i^*\}} n_{i'}\Bigg)} + \sum_{i'\in[k]}\frac{1}{2}\log{\left(a + n_{i'}\right)},
\end{align}
for any $i^{*}\in[k]$. Now, without loss of generality, assume that $i^*=1$, and consider the star graph $\mathcal{G}_{\mathrm{S}}$ with spokes emanating from the node corresponding to the first item with the following edge weights $n_{1j} = n_{j1} = 1$, for all $j\in[k]\setminus\{1, 2\}$, $n_{12}=n-(k-2)$, and $n_{ij} = 0$, otherwise. We claim that this configuration minimizes \eqref{eq:star-sum-rate} and we prove this claim by showing that any deviations will increase the value of \eqref{eq:star-sum-rate}. First, it is easy to see that amongst all possible edge weight assignments for star graphs with central node $i^*=1$, the edge weight assignment of $\mathcal{G}_{\mathrm{S}}$ minimizes the sum in  \eqref{eq:star-sum-rate} by the concavity of the logarithm function. Now, suppose that we shift part of the weight $n_{1j} > 0$ of an edge $(1, j)$, for $j\in[k]\setminus\{1\}$, to create a new edge $(j, i)$ with weight $n_{ji}$ such that $i\in[k]\setminus\{1\}$. Since we have
\begin{equation}
\displaystyle\frac{\partial S }{\partial n_{i}}= \frac{2n- \sum_{i'\in[k]\setminus\{i^{*}, i\}}n_{i'}}{\left(a+n_i\right)\left(a + 2n-\sum_{i'\in[k]\setminus\{i^{*}\}}n_{i'}\right)} > 0,
\end{equation}
for all $i\in[k]\setminus\{i^{*}\}$, we conclude that the sum in \eqref{eq:star-sum-rate} will be increased by the new configuration. Suppose instead that from the star graph configuration we shift part of the weight from the edge $(1, 2)$ with the most heavy weight $n_{12} >0$ to create a new edge $(j, i)$ with weight  $n_{ji}$ such that $i\in[k]\setminus\{1\}$ and $j\in[k]\setminus\{2\}$. We can  actually think of this transition as if it was done in two stages: At the first stage, we shift the weight $n_{12}$ from the edge $(1, 2)$ to the edge $(1, i)$ with weight $n_{1i}$, and at the second stage we shift the weight $n_{1i}$ from the edge $(1, i)$ to the edge $(j, i)$ with weight $n_{ji}$. But we know from the previous arguments that both stages of this transition will necessarily increase the sum in \eqref{eq:star-sum-rate}. Finally, we note that the types of deviations we considered are exhaustive, since for the graph to be connected, we must have $n_i >0$, for each $i\in[k]$, i.e., in any deviation we consider at least one element of each row of the adjacency matrix $\mathbf{N}$ of the graph must be non-zero.  So, the proof of the claim for the star graph is complete. 

Next we proceed with the proof of the claim concerning the chain graph. Note that any tree  has exactly $k-1$ non-zero edges with weights $n_{ij}$, for $(i, j)\in\mathcal{I}_o[k]$, and $n_i >0$, for all $i\in[k]$. To prove the extremality of the chain graph amongst trees, one can easily show that starting from the chain graph configuration, shifting any weight from any of the upper diagonal edges (in the adjacency matrix) into any position on its right (and similarly shifting the weights in the symmetrical positions of the matrix to preserve the overall symmetry) will result in an increase in the sum in~\eqref{eq:star-sum-rate}, and hence decrease in the lower bound on the Bayes risk. Similarly, removing any such weight entirely from the elements in the upper diagonal edges will decrease in the lower bound on the Bayes risk. This proves that the chain graph minimizes the lower bound on the Bayes risk in \eqref{eq:Bayes-Risk-LB-general} amongst all trees. 
\end{IEEEproof}

\bibliographystyle{IEEEtran}
\bibliography{LBBayesRiskBTLrefs}
\end{document}

\maketitle

This document contains some auxiliary lemmata for and proofs of propositions stated in  the paper ``Lower bounds on the Bayes risk of the Bayesian BTL model with applications to comparison graphs''.
\section*{Lemmas \ref{lem:prob-D-given-lambda}, \ref{lem:prob-lambda-D-Z}, and \ref{lem:p-L-given-DZ}}
\begin{lem}\label{lem:prob-D-given-lambda}
For the Bayesian BTL model introduced in Section II-A, the following conditional density holds:
\begin{equation}
p(\mathbf{W}|\boldsymbol{\lambda})  =  \prod_{(i,j)\in\mathcal{I}_o[k]}  \mathcal{B}(w_{ij}; n_{ij}, P_{ij}).
\end{equation}
\end{lem}
\begin{IEEEproof}
By the BTL model assumption, we can write
\begin{align}
p(\mathbf{W}|\boldsymbol{\lambda})  &= \prod_{(i,j)\in\mathcal{I}_o[k]}  {n_{ij} \choose w_{ij}} \prod_{(i,j)\in\mathcal{I}[k]}    P_{ij}^{w_{ij}}\\
 &= \prod_{(i,j)\in\mathcal{I}_o[k]}  {n_{ij} \choose w_{ij}}P_{ij}^{w_{ij}}   \prod_{(i,j)\in\mathcal{I}[k]\setminus\mathcal{I}_o[k]}P_{ij}^{w_{ij}} \\
  &= \prod_{(i,j)\in\mathcal{I}_o[k]}  {n_{ij} \choose w_{ij}}P_{ij}^{w_{ij}}   P_{ji}^{w_{ji}} \\
  &= \prod_{(i,j)\in\mathcal{I}_o[k]}  {n_{ij} \choose w_{ij}}P_{ij}^{w_{ij}}   \left(1-P_{ij}\right)^{n_{ij}-w_{ij}}.
\end{align}
\end{IEEEproof}

\begin{lem}\label{lem:prob-lambda-D-Z}
For the Bayesian BTL model introduced in Section II-A, the following joint density holds:
\begin{equation} \label{eq:joint-prob-simplified} 
 p(\boldsymbol{\lambda},\mathbf{W}, \mathbf{Z})  = \left(\prod_{i\in[k]} C(a_i, b_i)\right)  
 \left(\prod_{(i,j)\in\mathcal{I}_o[k]}  {n_{ij} \choose w_{ij}} \displaystyle\frac{ z_{ij}^{n_{ij}-1}}{\Gamma{(n_{ij})}}\right) 
 \left(\displaystyle\prod_{i\in[k]}   \lambda_i^{a_i+w_i-1}e^{-(b_i+z_i)\lambda_i}\right),
\end{equation}
where $w_i$ and $z_i$ are given by 
\begin{equation}\label{eq:wi}
w_i := \sum_{j\in[k]\setminus\{i\}} w_{ij}
\end{equation}
and 
\begin{equation}\label{eq:zi}
\zeta_i := \sum_{j\in[k]\setminus\{i\}} \zeta_{ij} 
\end{equation}
for all $i\in[k]$. 
\end{lem}
\begin{IEEEproof}
 We know that, by assumption, we have the following prior density:
\begin{equation}
p(\boldsymbol{\lambda})= \prod_{i=1}^{k} \mathcal{G}(\lambda_i: a_i, b_i) =  \prod_{i=1}^{k} C(a_i, b_i) \lambda_i^{a_i-1}e^{-b_i\lambda_i},
\end{equation}
where
 \begin{equation}
C(a_i, b_i) = \left(\displaystyle\frac{b_i^{a_i}}{\Gamma(a_i)}\right)^k.
 \end{equation}
We also know by \cite[Eq.~(2.1)]{CaronDoucet}  that 
\begin{equation}\label{eq:prob-Z-given-D-lambda}
Z_{ij}|\lambda_i, \lambda_j \sim p(\zeta_{ij}|\lambda_i, \lambda_j, n_{ij} )= \mathcal{G}(\zeta_{ij}; n_{ij}, \lambda_i+\lambda_j) 
\end{equation}
holds, for all $(i, j)\in\mathcal{I}[k]$. Using Lemma \ref{lem:prob-D-given-lambda},  the joint density $p(\boldsymbol{\lambda}, \mathbf{W}, \mathbf{Z})=p(\boldsymbol{\lambda})p(\mathbf{W}|\boldsymbol{\lambda})p(\mathbf{Z}|\mathbf{W}, \boldsymbol{\lambda})$ is obtained in \eqref{eq:prob-lambda-D-Z}  by re-arranging the terms of the product as follows:
\begin{align}
&p(\boldsymbol{\lambda},\mathbf{W}, \mathbf{Z})  \nonumber\\
&= \left(\prod_{1\leq i< j \leq k }   {n_{ij} \choose w_{ij}}\displaystyle\frac{\lambda_i^{w_{ij}}\lambda_j^{n_{ij}-w_{ij}}}{\left(\lambda_i+\lambda_j \right)^{n_{ij}}}\right) 
\left( \prod_{1\leq i< j \leq k : n_{ij}>0}  \displaystyle\frac{ (\lambda_i+\lambda_j)^{n_{ij}} z_{ij}^{n_{ij}-1}  e^{-(\lambda_i+\lambda_j)z_{ij}}}{\Gamma{(n_{ij})}}\right) \nonumber   \\
&\qquad \times \left(\prod_{i=1}^{k} C(a_i, b_i) \lambda_i^{a_i-1}e^{-b_i\lambda_i}\right) \\
 &=  \left(\prod_{i=1}^{k} C(a_i, b_i)\right) \left(\prod_{1\leq i< j \leq k : n_{ij}>0}   {n_{ij} \choose w_{ij}}\displaystyle\frac{ z_{ij}^{n_{ij}-1}}{\Gamma{(n_{ij})}}\right) \lambda_1^{a_1-1}  \left(\displaystyle\prod_{j=2}^{k}\lambda_1^{w_{1j}}\right)e^{-b_1\lambda_1} \left(\displaystyle\prod_{j=2}^{k} e^{-z_{1j}\lambda_1}\right) \nonumber\\
& \qquad\times \lambda_2^{n_{12}-w_{12}}e^{-\lambda_2z_{12}} \lambda_2^{a_2-1}  \left(\displaystyle\prod_{j=3}^{k}\lambda_2^{w_{2j}}\right)e^{-b_2\lambda_2} \left(\displaystyle\prod_{j=3}^{k} e^{-z_{2j}\lambda_2}\right) \nonumber\\
&\qquad\times \ldots \times \lambda_k^{n_{1k}-w_{1k}}e^{-\lambda_kz_{1k}} \lambda_k^{n_{2k}-w_{2k}}e^{-\lambda_kz_{2k}} \ldots \lambda_k^{n_{(k-1)k}-w_{(k-1)k}}e^{-\lambda_k z_{(k-1)k}} \lambda_k^{a_k-1}e^{-b_k\lambda_k}\\
&=    \left(\prod_{i=1}^{k} C(a_i, b_i)\right) \left(\prod_{1\leq i< j \leq k: n_{ij}>0}   {n_{ij} \choose w_{ij}}\displaystyle\frac{ z_{ij}^{n_{ij}-1}}{\Gamma{(n_{ij})}}\right) \lambda_1^{a-1}  \left(\displaystyle\prod_{j=2}^{k}\lambda_1^{w_{1j}}\right)e^{-b\lambda_1} \left(\displaystyle\prod_{j=2}^{k} e^{-z_{1j}\lambda_1}\right) \nonumber  \\
& \qquad\times \ldots \times\lambda_k^{w_{k1}}e^{-\lambda_k z_{k1}} \lambda_k^{w_{k2}}e^{-\lambda_k z_{k2}} \ldots \lambda_k^{w_{k(k-1)}}e^{-\lambda_k z_{k(k-1)}} \lambda_k^{a_k-1}e^{-b_k\lambda_k}\\
\label{eq:prob-lambda-D-Z}&=   \left(\prod_{i=1}^{k} C(a_i, b_i)\right) \left(\prod_{1\leq i< j \leq k : n_{ij}>0}  {n_{ij} \choose w_{ij}} \displaystyle\frac{ z_{ij}^{n_{ij}-1}}{\Gamma{(n_{ij})}}\right)  \left(\displaystyle\prod_{i=1}^{k}   \lambda_i^{a_i+w_i-1}e^{-(b_i+z_i)\lambda_i}\right).
\end{align}
\end{IEEEproof}

\begin{lem}\label{lem:p-L-given-DZ}
The variables of the Bayesian BTL model introduced in Section II-A obey the following conditional distribution: 
\begin{equation}\label{eq:p-L-given-DZ}
p(\boldsymbol{\lambda}|\mathbf{W},\mathbf{Z}) = \displaystyle\prod_{i\in[k]}\mathcal{G}(\lambda_i; a_i+w_i, b_i+z_i),
 \end{equation}
 where $w_i$ and $z_i$ are given by \eqref{eq:wi} and \eqref{eq:zi}, respectively. 
\end{lem}
\begin{IEEEproof}
Note that by definition we have $p(\boldsymbol{\lambda}|\mathbf{Z} \mathbf{W}) = p(\boldsymbol{\lambda},\mathbf{Z}, \mathbf{W}) /p(\mathbf{Z}, \mathbf{W}) $. We have already computed the joint density $p(\boldsymbol{\lambda}, \mathbf{W}, \mathbf{Z})$ in Lemma \ref{lem:prob-lambda-D-Z}.
Now, we evaluate 
\begin{equation}\label{eq:p-DZ}
p(\mathbf{W}, \mathbf{Z}) = \int_{\boldsymbol{\lambda}} p(\boldsymbol{\lambda}, \mathbf{W}, \mathbf{Z}) d\boldsymbol{\lambda}.
\end{equation}
Looking carefully at \eqref{eq:joint-prob-simplified}, one can easily see that \eqref{eq:p-DZ}  equals
\begin{equation}\label{eq:p-DZ-expression}
p(\mathbf{W},\mathbf{Z}) =   \left(\prod_{i\in[k]} C(a_i, b_i)\right)  
\left(\prod_{(i,j)\in\mathcal{I}_o[k]}   {n_{ij} \choose w_{ij}}\displaystyle\frac{ z_{ij}^{n_{ij}-1}}{\Gamma{(n_{ij})}}\right) 
\left(\displaystyle\prod_{i\in[k]} \Gamma(a_i+w_i)(b_i+z_i)^{-(a_i+w_i)}\right),
\end{equation}
where $w_i$ and $z_i$ are given by \eqref{eq:wi} and \eqref{eq:zi}, respectively. From \eqref{eq:joint-prob-simplified}  and \eqref{eq:p-DZ-expression}, we obtain
  \begin{equation}
p(\boldsymbol{\lambda}|\mathbf{W},\mathbf{Z}) = \displaystyle\prod_{i\in[k]}\mathcal{G}(\lambda_i; a_i+w_i, b_i+z_i).
 \end{equation}
\end{IEEEproof}

\section*{Lemma 4}
\begin{lem} \label{lem:asymp}
For the Bayesian BTL model introduced in Section II-A, the following holds:
\begin{equation}
\lim_{n_i\to\infty} \log{\left(1+O\left(\Expt\left[\frac{1}{a_i + \Omega_i }\right]\right)\right)} = 0.
\end{equation}
\end{lem}
\begin{IEEEproof}
Let us first observe that, for any fixed $i\in[k]$, we have  $\Expt\left[\Omega_i\right] = \Expt\left[\sum_{j\in[k]\setminus\{i\}}\Expt\left[\Omega_{ij}|\Lambda_i, \Lambda_j\right]\right]$ and $p(\Omega_{ij}|\lambda_i, \lambda_i) = \mathcal{B}(w_{ij}; n_{ij}, P_{ij})$. Thus, the probability (or moment) generating function of the random variable $\Omega_i$ conditional on $\boldsymbol{\Lambda} = \boldsymbol{\lambda} $ is given by \cite{pgf-binomial}
\begin{equation}
\Pi_{\Omega_i|\boldsymbol{\Lambda} = \boldsymbol{\lambda}}(s) = \prod_{j\in[k]\setminus\{i\}}\left((1-P_{ij}) + P_{ij}s)\right)^{n_{ij}} = \exp  \left\{\sum_{j\in[k]\setminus\{i\}}n_{ij}\ln\left((1-P_{ij}) + P_{ij}s)\right) \right\},
\end{equation}
where $\ln$ stands for the natural logarithm function. Furthermore, one can write
\begin{equation}
\Expt\left[\frac{1}{a_i + \Omega_i }\right] = \int_{0}^{1} \exp\{(a_i - 1)\ln s\} \Pi_{\Lambda_i|\boldsymbol{\Lambda} = \boldsymbol{\lambda}}(s)  ds.
\end{equation}
Now, since 
\begin{equation}
\lim_{n_i\to\infty}\exp\{(a_i - 1)\ln s\} \Pi_{\Lambda_i|\boldsymbol{\Lambda} = \boldsymbol{\lambda}}(s) = 0,
\end{equation} 
and $\exp\{(a_i - 1)\ln s\} \Pi_{\Lambda_i|\boldsymbol{\Lambda} = \boldsymbol{\lambda}}(s) \leq \exp\{(a_i - 1)\ln s\}$ holds, for any $s\in(0, 1)$, we conclude by the dominated convergence theorem that 
\begin{equation}
\lim_{n_i\to\infty} \Expt\left[\frac{1}{a_i + \Omega_i }\right]  =  \int_{0}^{1} \lim_{n_i\to\infty}\exp\{(a_i - 1)\ln s\} \Pi_{\Lambda_i|\boldsymbol{\Lambda} = \boldsymbol{\lambda}}(s) ds = 0.
\end{equation} 
This concludes the proof.
\end{IEEEproof}

\section*{Proof of Proposition 2}
\begin{IEEEproof}[Proof of Proposition 2]
Consider the BTL model with home-field advantage introduced in Section V. We first note that the following relation holds for the defined variables: 
  \begin{align}
I(\boldsymbol{\Lambda},\Theta;  \boldsymbol{\Omega}^{\mathrm{h}},\mathbf{Z}^{\mathrm{h}}) - h(\boldsymbol{\Lambda}, \Theta) 
&= \Expt\left[\log{p(\boldsymbol{\Lambda},\Theta| \boldsymbol{\Omega}^{\mathrm{h}},\mathbf{Z}^{\mathrm{h}})}\right]\\
&\leq  \Expt\left[\log{p(\boldsymbol{\Lambda}| \boldsymbol{\Omega}^{\mathrm{h}},\mathbf{Z}^{\mathrm{h}},\Theta)}\right]\\
\label{eq:exponent-HA}&=\displaystyle\sum_{i\in[k]}\Expt\left[\log p(\Lambda_i| \mathbf{\Omega}^{\mathrm{h}}, \mathbf{Z}^{\mathrm{h}}, \Theta)\right]
  \end{align}
where  the conditional density of  the skill parameters of the model is as given by  \cite[Eq.~(17)]{CaronDoucet} 
\begin{multline}\label{eq:lambda-given-D-Z-theta}
\Lambda_i| \mathbf{W}^{\mathrm{h}} , \boldsymbol{\zeta}^{\mathrm{h}} , \theta \sim p(\lambda_i| \mathbf{W}^{\mathrm{h}}, \boldsymbol{\zeta}^{\mathrm{h}}, \theta) \\
= \mathcal{G}\left(\lambda_i; a_i+\displaystyle\sum_{j\in[k]\setminus\{i\}}w_{ij}^{\mathrm{h}}+\displaystyle\sum_{j\in[k]\setminus\{i\}}\left(n_{ji}^{\mathrm{h}}-w_{ji}^{\mathrm{h}}\right), 
 b_i+ \theta \displaystyle\sum_{j\in[k]\setminus\{i\}}\zeta_{ij}^{\mathrm{h}}  +  \displaystyle\sum_{j\in[k]\setminus\{i\}}\zeta_{ji}^{\mathrm{h}} \right),
\end{multline}
for any $i\in[k]$.  Now similar to the proof of Proposition 1, we want to get an asymptotic upper bound on \eqref{eq:exponent-HA} by using Proposition 3 given at the end of Appendix A.  For that purpose, we first claim that
\begin{equation}
\lim_{n_i \to \infty}\log\left(1+ \Expt\left[\left(a_i+\displaystyle\sum_{j\in[k]\setminus\{i\}}\Omega_{ij}^{\mathrm{h}}+\displaystyle\sum_{j\in[k]\setminus\{i\}}\left(n_{ji}^{\mathrm{h}}-\Omega_{ji}^{\mathrm{h}}\right) \right)^{-1}\right]\right) = 0
\end{equation} 
holds. The result can be verified using similar steps to the proof of Lemma \ref{lem:asymp} stated in the previous section of this Supplementary Material. Thus, we can apply  Proposition 3 using the identifications $M\leftarrow \Lambda_i$, $A \leftarrow  a_i+\sum_{j\in[k]\setminus\{i\}}w_{ij}^{\mathrm{h}}+\sum_{j\in[k]\setminus\{i\}}\left(n_{ji}^{\mathrm{h}}-w_{ji}^{\mathrm{h}}\right)$, and $B\leftarrow   b_i+ \theta \sum_{j\in[k]\setminus\{i\}}\zeta_{ij}^{\mathrm{h}}  +  \sum_{j\in[k]\setminus\{i\}}\zeta_{ji}^{\mathrm{h}}$. It only remains to compute the expectations arising from the application of Proposition  3. We start by computing
%  where 
% \begin{equation}\label{eq:ci}
% c_i = a + \displaystyle\sum_{j\neq i} w_{ij}= a_i + w_i,
%  \end{equation}
%  \begin{equation}\label{eq:di}
%d_i = b + \displaystyle\sum_{j=1}^{i-1} z_{ji} + \displaystyle\sum_{j=i+1}^{K} z_{ij} = b_i + \displaystyle\sum_{j \neq i} z_{ij} 
% \end{equation}
\begin{align}
&\Expt\left[a_i+\displaystyle\sum_{j\in[k]\setminus\{i\}}w_{ij}^{\mathrm{h}}+\displaystyle\sum_{j\in[k]\setminus\{i\}}\left(n_{ji}^{\mathrm{h}}-w_{ji}^{\mathrm{h}}\right)\right] \nonumber\\
\label{eq:HA-t1}&= a_i+ \displaystyle\sum_{j\in[k]\setminus\{i\}}\Expt\left[\displaystyle\frac{n_{ij}^{\mathrm{h}}\Theta\Lambda_i}{\Theta\Lambda_i+\Lambda_j} \right] +   \displaystyle\sum_{j\in[k]\setminus\{i\}}\Expt\left[\displaystyle\frac{n_{ji}^{\mathrm{h}}\Lambda_i}{\Lambda_i+\Theta\Lambda_j} \right]
\end{align}
 where \eqref{eq:HA-t1} follows from  $\Omega_{ij}^{\mathrm{h}}\sim \mathcal{B}(w_{ij}^{\mathrm{h}}; n_{ij}^{\mathrm{h}}, Q_{ij})$ and 
$\left(n_{ji}^{\mathrm{h}}-\Omega_{ji}^{\mathrm{h}}\right)\sim\mathcal{B}(n_{ji}^{\mathrm{h}}-w_{ji}^{\mathrm{h}}; n_{ji}^{\mathrm{h}}, \overline{Q}_{ij})$. Then, we compute
  \begin{align}
&\Expt\left[ \left(b_i+ \Theta \displaystyle\sum_{j\in[k]\setminus\{i\}}Z_{ij}^{\mathrm{h}} +  \displaystyle\sum_{j\in[k]\setminus\{i\}}Z_{ji}^{\mathrm{h}}\right)\Lambda_i\right]   \nonumber\\
\label{eq:HA-t2}& = a_i + \displaystyle\sum_{j\in[k]\setminus\{i\}}\Expt\left[\displaystyle\frac{ n_{ij}^{\mathrm{h}}\Theta\Lambda_i}{\Theta\Lambda_i+\Lambda_j}\right]  +  \displaystyle\sum_{j\in[k]\setminus\{i\}} \Expt\left[\displaystyle\frac{n_{ji}^{\mathrm{h}} \Lambda_i}{\Lambda_i+\Theta\Lambda_j}\right]  
 \end{align}
 where \eqref{eq:HA-t2} follows from $Z_{ji}^{\mathrm{h}} |\lambda_i,\lambda_j, \theta \sim \mathcal{G}(\zeta_{ij}^{\mathrm{h}}; n_{ij}^{\mathrm{h}}, \theta\lambda_i+\lambda_j)$ and $\Lambda_i\sim\mathcal{G}(\lambda_i, a_i, b_i)$. Thus, as in the basic BTL model, the difference of the terms in \eqref{eq:HA-t1} and \eqref{eq:HA-t2} is zero.  Next, we note that the term $\Expt\left[\log{\Lambda_i}\right] = \psi(a_i)-\log{b_i}$ remains unchanged, and the final term equals
\begin{align}
&\frac{1}{2}\log{\left(\Expt\left[a_i+\displaystyle\sum_{j\in[k]\setminus\{i\}}w_{ij}^{\mathrm{h}}+\displaystyle\sum_{j\in[k]\setminus\{i\}}\left(n_{ji}^{\mathrm{h}}-w_{ji}^{\mathrm{h}}\right)\right]\right)} \nonumber \\
&=\frac{1}{2}\log{\left(a_i + \displaystyle\sum_{j\in[k]\setminus\{i\}}F_{ij}(n_{ij}^{\mathrm{h}}, n_{ji}^{\mathrm{h}}, a_i,b_i, p(\Theta))
\right)},
\end{align} 
where
\begin{equation}\label{eq:HA-theta-function}
F_{ij}(n_{ij}^{\mathrm{h}}, n_{ji}^{\mathrm{h}}, a_i,b_i, p(\Theta)) 
=  \Expt\left[\displaystyle\frac{\Theta\Lambda_i}{\Theta\Lambda_i+\Lambda_j}\right]n_{ij}^{\mathrm{h}} + \Expt\left[\displaystyle\frac{\Lambda_i}{\Lambda_i+\Theta\Lambda_j}\right]n_{ji}^{\mathrm{h}}  ,
\end{equation}
for any $(i, j)\in\mathcal{I}[k]$. Thus, we obtain
\begin{multline}
\frac{1}{k}\left(I(\boldsymbol{\Lambda}; \boldsymbol{\Omega}^{\mathrm{h}}\mathbf{Z}^{\mathrm{h}}) - h(\boldsymbol{\Lambda})\right)\\
\lesssim_{n_i} \frac{1}{k}\displaystyle\sum_{i\in [k]} \vast( - \frac{1}{2}\log{(2\pi)} + \log{b_i}-\psi(a_i) 
+\frac{1}{2}\log{\left(a_i + \displaystyle\sum_{j\in[k]\setminus\{i\}}F_{ij}(n_{ij}^{\mathrm{h}}, n_{ji}^{\mathrm{h}}, a_i,b_i, p(\Theta))
\right)}\vast).
\end{multline}
This concludes the proof.
\end{IEEEproof}

\section*{Proof of Theorem 5}
\begin{IEEEproof}
 Using the likelihood function in (43), we obtain the  log-likelihood as follows:
\begin{align}
&\log p(\matD^{\mathrm{h}}, \bmlambda| \theta) \nonumber\\
&= \sum_{(i,j) \in \mathcal{I}[k]} \left[\log {n_{ij}^{\mathrm{h}} \choose w^{\mathrm{h}}_{ij}}+ {w^{\mathrm{h}}_{ij}} \log \theta \lambda_i - w^{\mathrm{h}}_{ij} \log (\theta \lambda_i+\lambda_j)  + {(n_{ij}^{\mathrm{h}} - w^{\mathrm{h}}_{ij})} \log \lambda_j - {(n_{ij}^{\mathrm{h}} - w^{\mathrm{h}}_{ij})} \log (\theta \lambda_i + \lambda_j)\right] \nonumber\\
&\qquad \qquad+ \sum_{i = 1}^k a_i \log b - \log \Gamma(a_i) 
+ (a_i-1) \log \lambda_i -b\lambda_i 
\label{eq:LogLikelihoodHCRBHA}
\end{align}
 
Differentiating \eqref{eq:LogLikelihoodHCRBHA} w.r.t.\ $\lambda_i$, we obtain
\begin{equation}
\frac{\partial \log p(\matD^{\mathrm{h}}, \lambda_i| \theta)}{\partial \lambda_i} = \frac{(a_i-1 )+ \sum_{j = 1}^k (w^{\mathrm{h}}_{ij} +(n_{ji}^{\mathrm{h}} - w^{\mathrm{h}}_{ji}))}{\lambda_i} - \sum_{j = 1}^k \frac{n_{ij}^{\mathrm{h}} \theta}{\theta\lambda_i + \lambda_j} - \sum_{j = 1}^k \frac{n_{ji}^{\mathrm{h}} }{\theta\lambda_j + \lambda_i} - b
\label{diff1lambdaiHA}
\end{equation}
Differentiating the above w.r.t.\ $\lambda_i$ again, we obtain
\begin{equation}
\label{eq:diff2iiHA}
\frac{\partial^2 \log p(\matD^{\mathrm{h}}, \lambda_i|\theta)}{\partial \lambda_i^2} = \frac{-(a_i-1 )- \sum_{j = 1}^k (w^{\mathrm{h}}_{ij}+(n_{ji}^{\mathrm{h}} - w^{\mathrm{h}}_{ji}))}{\lambda_i^2} + \sum_{j = 1}^k \frac{n_{ij}^{\mathrm{h}}  \theta^2}{(\theta\lambda_i + \lambda_j)^2} + \sum_{j = 1}^k \frac{n_{ji}^{\mathrm{h}}}{(\theta\lambda_i + \lambda_j)^2}
\end{equation}
Hence, we obtain the diagonal entries of $\matH^{\bmLambda}$ is given by $[\matH^{\bmLambda}]_{i,i}$ as 
\begin{align}
 [\matH^{\bmLambda}]_{i,i} & = -\Ex\left[ \frac{\partial^2 \log p(\bmDelta^{\mathrm{h}},\Lambda_i|\theta)}{\partial \Lambda_i^2} \right] \\
&= \Ex\left[ \frac{(a_i-1)}{\Lambda_i^2}+ \theta \sum_{j = 1}^k \frac{ n_{ij}^{\mathrm{h}}\Lambda_j }{\Lambda_i(\theta\Lambda_i + \Lambda_j)} + \theta \sum_{j = 1}^k \frac{ n_{ji}^{\mathrm{h}}\Lambda_j }{\Lambda_i(\theta\Lambda_j + \Lambda_i)}\right]
\end{align}
Furthermore, differentiating \eqref{diff1lambdaiHA} w.r.t.\ $\lambda_j$, we obtain
\begin{equation}
\label{eq:diff2ijHA}
\frac{\partial^2 \log p(\matD^{\mathrm{h}}, \lambda_i|\theta)}{\partial \lambda_i \partial \lambda_j} =   \frac{ \theta n_{ij}^{\mathrm{h}} }{(\theta\lambda_i + \lambda_j)^2} +   \frac{ \theta n_{ji}^{\mathrm{h}} }{(\theta\lambda_j + \lambda_i)^2}
\end{equation}
Hence, we obtain the off-diagonal entries of $\matH^{\bmLambda}$ given by $[\matH^{\bmLambda}]_{i,j}$ as 
\begin{align}
&[\matH^{\bmLambda}]_{i,j} = -\Ex\left[ \frac{\partial^2 \log p(\bmDelta^{\mathrm{h}},\Lambda_i| \theta)}{\partial \Lambda_i \partial \Lambda_j}\right] = - \frac{ \theta n_{ij}^{\mathrm{h}}}{(\theta\Lambda_i + \Lambda_j)^2} - \frac{ \theta n_{ji}^{\mathrm{h}} }{(\theta\Lambda_j + \Lambda_i)^2}.
\end{align}
From the above, we see that for $\theta = 1$, $n^{\mathrm{h}}_{ij} = n^{\mathrm{h}}_{ji}$, \eqref{eq:diff2iiHA} and \eqref{eq:diff2ijHA} is equal to (81) and (82), respectively. Hence, HIM is same as BIM for $\theta = 1$. 
Differentiating \eqref{eq:LogLikelihoodHCRBHA} twice w.r.t.\ $\theta$ we obtain
\begin{align}
&\frac{\partial^2 \log p(\matD^{\mathrm{h}}, \bmlambda|\theta)}{\partial \theta^2} = -\frac{\sum_{i = 1}^k \sum_{j = 1}^k w^{\mathrm{h}}_{ij}}{\theta^2} + \sum_{i = 1}^k \sum_{j = 1}^k\frac{  n^{\mathrm{h}}_{ij}\lambda_i^2}{(\theta \lambda_i+\lambda_j)^2}
\end{align}
The above expression allows us to obtain $[\matH^{\theta}]_{1,1}$ given by
\begin{align}
 [\matH^{\theta}]_{1,1}  &= \Ex\left[ \frac{\sum_{i = 1}^k \sum_{j = 1}^k w^{\mathrm{h}}_{ij}}{\theta^2} - \sum_{i = 1}^k \sum_{j = 1}^k\frac{\Lambda_i^2  n^{\mathrm{h}}_{ij}}{(\theta \Lambda_i+\Lambda_j)^2}\right]\\
&=\sum_{i = 1}^k \sum_{j = 1}^k\frac{ n^{\mathrm{h}}_{ij}\Lambda_i}{\theta(\theta\Lambda_i+\Lambda_j)} -  \sum_{i = 1}^k \sum_{j = 1}^k\frac{\Lambda_i^2 n^{\mathrm{h}}_{ij}}{(\theta \Lambda_i+\Lambda_j)^2}.
\end{align}
Furthermore, differentiating \eqref{diff1lambdaiHA}  w.r.t.\ $\theta$, we obtain
\begin{equation}
\frac{\partial^2 \log p(\matD^{\mathrm{h}}, \bmlambda, \theta)}{\partial \lambda_i \partial \theta} = - \sum_{j = 1}^k \frac{n_{ij}^{\mathrm{h}} }{\theta\lambda_i + \lambda_j} + \sum_{j = 1}^k \frac{n_{ij}^{\mathrm{h}}\theta\lambda_i }{(\theta\lambda_i + \lambda_j)^2} + \sum_{j = 1}^k \frac{n_{ji}^{\mathrm{h}} \lambda_j}{(\theta\lambda_j + \lambda_i)^2} 
\end{equation}
The above expression allows us to obtain $[\matH^{\bmLambda,\theta}]_{i,1}$ given by
\begin{equation}
[\matH^{\bmLambda,\theta}]_{i,1} = \sum_{j = 1}^k \frac{n_{ij}^{\mathrm{h}} }{\theta\Lambda_i + \Lambda_j} - \sum_{j = 1}^k \frac{n_{ij}^{\mathrm{h}}\theta\Lambda_i }{(\theta\Lambda_i + \lambda_j)^2} - \sum_{j = 1}^k \frac{n_{ji}^{\mathrm{h}} \Lambda_j}{(\theta\Lambda_j + \Lambda_i)^2}.
\end{equation}
Using the above given results, we obtain the HIM for HCRB as
\begin{align}
&\matH^{\bmLambda,\theta} :=
\begin{bmatrix}
\mathbf{H}^{\bmLambda} &   \mathbf{H}^{\bmLambda,\theta}\\
(\mathbf{H}^{\bmLambda,\theta})^T & \mathbf{H}^{\theta}
\end{bmatrix}
\end{align}
where, $[{\matH}^{\bmLambda}]_{i,i} $, $[\matH^{\bmLambda}]_{i,j} $, $[\matH^{\theta}]_{1,1}$, and $[\matH^{\bmLambda,\theta}]_{i,1} $  are computed using \eqref{lem:Expt-HCRB} and \eqref{lem:Expt-SecondMoment} and are as in (47), (48), (49), and (50), respectively. 
\end{IEEEproof} 

\begin{figure}[htp]
\begin{center}
        \includegraphics[scale=0.5]{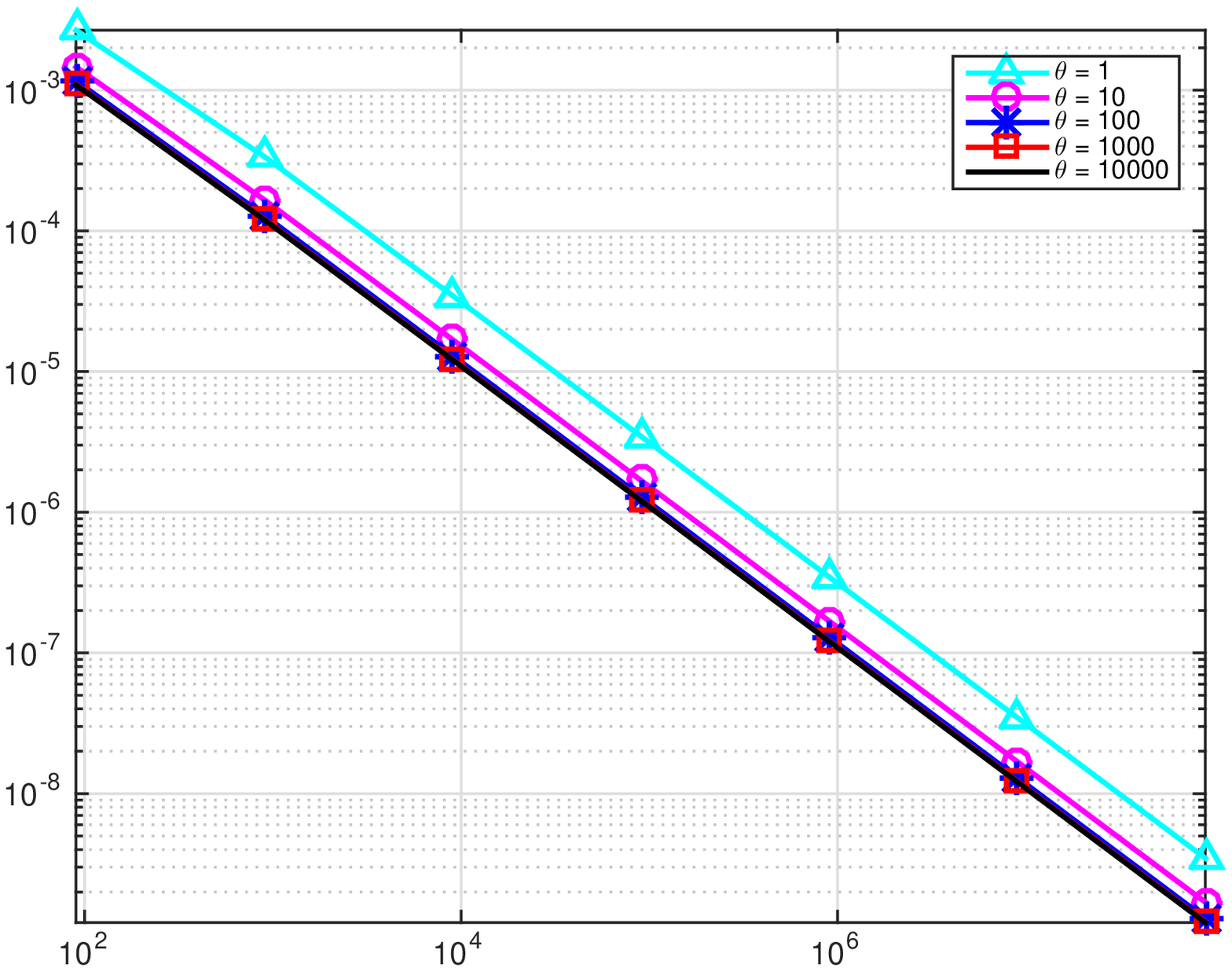}         
                  \caption{The HCRB in the presence of the home-field advantage parameter $\theta>1$ as derived in  Theorem 5 for squared $L^2$  norm. The parameters of the prior distribution in (9) are chosen as $a=5$ and $b=ak-1$ with $n_{ij}^{\mathrm{h}} = n_{ij}$, for $(i, j)\in\mathcal{I}_o[k]$, and for $k=10$ items. }
\label{BCRBHA}
\end{center}
\end{figure}

In Fig.~\ref{BCRBHA}, we illustrate the HCRB with home-field advantage parameter $\theta$. 

\section*{Lemmas \ref{lem:Expt-HCRB} and \ref{lem:Expt-SecondMoment}}
\begin{lem}\label{lem:Expt-HCRB}
Let $\Lambda_i$, $\Lambda_j$ be two non-negative random variables distributed according to a Gamma distribution given by  $\mathcal{G}(\Lambda_i; a_i, b)$ and $\mathcal{G}(\Lambda_j; a_j, b)$, where $a_i, a_j$ and $b$ determines respectively the shape and rate parameters of a random gamma distribution of $\Lambda_i$ and $\Lambda_j$. Then,  
\begin{align}
\label{eq:exptLem4}
&\Expt\left[ \frac{\Lambda_i}{\theta\Lambda_i+\Lambda_j}\right] = (-\theta)^{a_j - 1}B(a_i,a_j)\nonumber\\
& \left[\frac{\theta^{-(a_i+a_j)}}{(a_i+a_j)} {}_2F_1\left(a_i+a_j,a_i+a_j;a_i+a_j+1,\frac{(\theta-1)}{\theta}\right)+\sum_{k' = 1}^{a_j-1}(-\theta)^{-k'}(k'-1)!\frac{\Gamma(a_i+a_j-k')}{\Gamma(a_i+a_j)}\right],
\end{align}
where ${}_2F_1(\cdot)$ is the Hypergeometric function.
\end{lem}
\begin{IEEEproof}
The expectation of $\frac{\lambda_i}{(\theta\lambda_i + \lambda_j)}$ can be computed as
\begin{align}
&\Expt\left[ \frac{\lambda_i}{(\theta\lambda_i + \lambda_j)}\right] =c_{\bmlambda}\int_{\lambda_i = 0}^\infty \int_{\lambda_j = 0}^{\infty} \frac{\lambda_i}{(\theta\lambda_i + \lambda_j)} \lambda_i^{a_i - 1} e^{-b\lambda_i} \lambda_j^{a_j - 1} e^{-b\lambda_j} d\lambda_i d\lambda_j,
\end{align}
where $c_{\bmlambda} = \frac{b^{(a_i+a_j)}}{\Gamma(a_i)\Gamma(a_j)}$. We simplify the inner integral (w.r.t. $\lambda_j$) using the relation given by \cite{gradshteyntable}
\begin{align}
&\int_0^\infty \frac{x^n e^{-\mu x}}{x+\beta} dx = (-1)^{n-1}\beta^n e^{\beta\mu}\textnormal{Ei}(-\beta\mu)+\sum_{k = 1}^n (k-1)! (-\beta)^{(n-k)}\mu^{-k},
\end{align}
for $\lvert \arg(\beta)\rvert < \pi$ and $\Re(\mu) > 0$, where $\textnormal{Ei}(\cdot)$ is the exponential integral. Furthermore, we solve the resulting expression using the following relation:
\begin{align}
&\int_{0}^\infty x^p e^{(ax)} \textnormal{E}_1(bx) dx = \frac{\Gamma(p+1)}{p+1}\frac{1}{b^{(p+1)}}{}_2F_1(p+1,p+1;p+2,a/b),
\end{align}
where we have used the fact that an alternate form of the exponential integral is given by $\textnormal{E}_1(x) = -\textnormal{Ei}(-x)$, and  $a > b$ and $p > -1$. Hence, we obtain the given expression for $ \Expt\left[ \frac{\Lambda_i}{\theta\Lambda_i+\Lambda_j}\right] $. 
\end{IEEEproof}
Further, we generalize \eqref{eq:exptLem4} for $ t_i,t_j \in (-\infty,\infty)$ as 
\begin{align}
&\mu_{\Lambda_i\Lambda_j} (t_i,t_j,\theta) := \frac{b^{-(t_i+t_j)}(-\theta)^{a'_j - 1}}{\Gamma(a_i)\Gamma(a_j)}\\
& \left[\frac{\Gamma(a'_i + a'_j)\theta^{-(a'_i+a'_j)}}{(a'_i+a'_j)} {}_2F_1\left(a'_i+a'_j,a'_i+a'_j;a'_i+a'_j+1,\frac{(\theta-1)}{\theta}\right)+\sum_{k' = 1}^{a'_j-1}(-\theta)^{-k'}(k'-1)!\Gamma(a'_i+a'_j-k')\right]\nonumber,
\end{align}
where $a'_i = a_i+t_i$ and $a'_j = a_j+t_j$.

%\begin{cor}\label{cor:lem5Gengamma}
%Let $\Lambda_i$, $\Lambda_j$ be two non-negative random variables distributed according to a Gamma distribution given by  $\mathcal{G}(\Lambda_i; a_i', b)$ and $\mathcal{G}(\Lambda_j; a_j', b)$, where $a'_i, a'_j$ and $b$ determines respectively the shape and rate parameters of a random gamma distribution of $\Lambda_i$ and $\Lambda_j$. Then,  
%\begin{align}
%&\mu'_{\Lambda_i\Lambda_j} (a_i,a_j,b,\theta)\triangleq \Expt\left[ \frac{\Lambda_i}{\theta\Lambda_i+\Lambda_j}\right] = (-\theta)^{a_j - 1}B(a_i,a_j)\nonumber\\
%& \left[\frac{\theta^{-(a_i+a_j)}}{(a_i+a_j)} {}_2F_1\left(a_i+a_j,a_i+a_j;a_i+a_j+1,\frac{(\theta-1)}{\theta}\right)+\sum_{k' = 1}^{a_j-1}(-\theta)^{-k'}(k'-1)!\frac{\Gamma(a_i+a_j-k')}{\Gamma(a_i+a_j)}\right],
%\end{align}
%where ${}_2F_1(\cdot)$ is the Hypergeometric function.
%\end{cor}

\begin{lem}\label{lem:Expt-SecondMoment}
Let $\Lambda_i$, $\Lambda_j$ be two non-negative random variables distributed according to a Gamma distribution given by  $\mathcal{G}(\Lambda_i; a_i, b)$ and $\mathcal{G}(\Lambda_j; a_j, b)$, where $a_i, a_j$ and $b$ determines respectively the shape and rate parameters of a random gamma distribution of $\Lambda_i$ and $\Lambda_j$. Then,  
\begin{align}
 \nu_{\Lambda_i,\Lambda_j}(t_i,t_j,\theta)& := \Expt\left[ \frac{1}{(\theta\Lambda_i+\Lambda_j)^2}\right] = (a_j-1)\mu_{\Lambda_i,\Lambda_j}(-1,-1,\theta) - b \mu_{\Lambda_i,\Lambda_j}(-1,0,\theta).
\end{align}
\end{lem}
\begin{IEEEproof} 
The expression $\Expt\left[ \frac{1}{(\theta\Lambda_i+\Lambda_j)^2}\right] $ is given by
\begin{align}
&\Expt\left[ \frac{1}{(\theta\Lambda_i+\Lambda_j)^2}\right] =  c_{\bmlambda}\int_{\lambda_i = 0}^\infty \int_{\lambda_j = 0}^{\infty} \frac{1}{(\theta\lambda_i + \lambda_j)^2} \lambda_i^{a_i - 1} e^{-b\lambda_i} \lambda_j^{a_j - 1} e^{-b\lambda_j} d\lambda_i d\lambda_j
\end{align}
Using integration by parts, the above expression can be written as
\begin{align}
 \Expt\left[ \frac{1}{(\theta\Lambda_i+\Lambda_j)^2}\right] = c_{\bmlambda}\int_{\lambda_i = 0}^\infty \int_{\lambda_j = 0}^{\infty}  \frac{((a_j-1)\lambda_j^{(a_j-2)}e^{-b\lambda_j} - b \lambda_j^{(a_j-1)}e^{-b\lambda_j})}{(\theta\lambda_i + \lambda_j)} \lambda_i^{a_i - 1} e^{-b\lambda_i}  d\lambda_j d\lambda_i
\end{align}
Now, we use Lemma \ref{lem:Expt-HCRB} to obtain the expression for $\nu_{\Lambda_i\Lambda_j}(a_i,a_j,b,\theta)$.
\end{IEEEproof}

\bibliographystyle{IEEEtran}
\bibliography{LBBayesRiskBTLrefs}